\documentclass[graybox]{svmult}

\usepackage{type1cm}        % activate if the above 3 fonts are
\usepackage{makeidx}         % allows index generation
\usepackage{graphicx}        % standard LaTeX graphics tool
\usepackage{multicol}        % used for the two-column index
\usepackage[bottom]{footmisc}% places footnotes at page bottom

\graphicspath{{figures/}}
\usepackage{subfig}
\usepackage{appendix}

\usepackage{newtxtext}       % 
\usepackage{newtxmath}       % selects Times Roman as basic font

\makeindex             % used for the subject index
\newif\ifcolouredsymbols

\colouredsymbolstrue
\colouredsymbolsfalse

\ifcolouredsymbols

  \newcommand{\newmathcommand}[3][0]{
    \newcommand{#2}[#1]{\ensuremath{{\mathcolour#3}}}}
  
\else

  \newcommand{\newmathcommand}[3][0]{
    \newcommand{#2}[#1]{\ensuremath{#3}}}
  
\fi
\newmathcommand{\Div}{\operatorname{div}}
\newmathcommand{\ones}{1}
\newcommand{\argmin}[1]{\underset{#1}{\operatorname{arg}\,\operatorname{min}}\;}

\newcommand{\Dc}{\mathcal{D}}
\newcommand{\Xc}{\mathcal{X}}

\newcommand{\Kc}{\mathcal{K}}
\newcommand{\Lc}{\mathcal{L}}
\newcommand{\Nc}{\mathcal{N}}
\newcommand{\Rc}{\mathcal{R}}
\newcommand{\Uc}{\mathcal{U}}
\newcommand{\Vc}{\mathcal{V}}

\newmathcommand{\R}{\mathbb{R}}
\newmathcommand{\N}{\mathbb{N}}

\newmathcommand{\opA}{K}
\newmathcommand{\imageA}{u}
\newmathcommand{\imageB}{v}
\newmathcommand{\dataA}{f}
\newmathcommand{\dataAnoisy}{\dataA^\delta}

\newmathcommand{\dataB}{g}

\newmathcommand{\spaceA}{\Uc}
\newmathcommand{\spaceB}{\Vc}
\newmathcommand{\spaceC}{\mathcal Z}

\newmathcommand{\subspaceA}{\Xc}

\newmathcommand[1]{\norm}{\|#1\|}
\newmathcommand[1]{\normA}{\norm{#1}_{\spaceA}}
\newmathcommand[1]{\normB}{\norm{#1}_{\spaceB}}
\newmathcommand[1]{\normC}{\norm{#1}_{\spaceC}}
\newmathcommand[1]{\Norm}{\left\|#1\right\|}
\newmathcommand[1]{\NormA}{\Norm{#1}_{\spaceA}}
\newmathcommand[1]{\NormB}{\Norm{#1}_{\spaceB}}
\newmathcommand[1]{\NormC}{\Norm{#1}_{\spaceC}}
\newmathcommand[2]{\innerp}{\langle#1,#2\rangle}
\newmathcommand[2]{\innerpA}{\innerp{#1}{#2}_{\spaceA}}
\newmathcommand[2]{\innerpB}{\innerp{#1}{#2}_{\spaceB}}

\newmathcommand{\linsymbol}{\Lc}

\newmathcommand{\comsymbol}{\Kc}

\newmathcommand{\adjsymbol}{\ast}

\newmathcommand{\rangesymbol}{\Rc}

\newmathcommand{\kernelsymbol}{\Nc}

\newmathcommand{\domainsymbol}{\Dc}

\newmathcommand{\orthsymbol}{\perp}

\newmathcommand[1]{\proj}{P_{#1}}

\newmathcommand{\minnormsymbol}{\dagger}

\newmathcommand{\RI}{\R_\infty}
\newmathcommand{\FunA}{E}
\newmathcommand{\FunB}{F}

\newmathcommand{\C}{\mathbb C}
\newmathcommand{\K}{\mathbb K}

\newmathcommand{\DataTerm}{D}

\newmathcommand{\topA}{\tau_\spaceA}
\newmathcommand{\topB}{\tau_\spaceB}
\newmathcommand{\Reg}{J}
\newmathcommand{\ConvTopA}{\overset{\topA}{\rightarrow}}

\newmathcommand{\VarReg}{\Phi}
\newmathcommand{\TV}{\operatorname{TV}}
\newmathcommand{\BV}{\operatorname{BV}}

\newmathcommand{\HilA}{\mathcal H}

\newmathcommand{\setA}{\mathcal C}
\newmathcommand{\ConvSetA}{\mathcal C}

\input{tikz}
\begin{document}

\title*{Joint phase reconstruction and magnitude segmentation from velocity-encoded MRI data}
\titlerunning{Joint phase reconstruction and magnitude segmentation}% for an abbreviated version of
\author{Veronica Corona, Martin Benning, Lynn F. Gladden, Andi Reci, Andrew J. Sederman, Carola-Bibiane Sch{\"o}nlieb}
% Use 
\authorrunning{Corona et al.}% for an abbreviated version of
\institute{Veronica Corona, Carola-Bibiane Sch{\"o}nlieb \at Department of Applied Mathematics and Theoretical Physics, University of Cambridge, UK, \email{\{vc324, cbs31\}@cam.ac.uk}
\and Martin Benning \at School of Mathematical Sciences, Queen Mary University of London, UK \email{m.benning@qmul.ac.uk}
\and  Lynn F. Gladden, Andi Reci, Andrew J. Sederman \at Department of Chemical Engineering and Biotechnology, University of Cambridge, UK \email{\{lfg1, ar622, ajs40\}@cam.ac.uk}}
%		\vspace{4mm}?"
%	{\footnotesize{$^1$ Department of Applied Mathematics and Theoretical Physics, University of Cambridge, UK\\
%	$^2$ School of Mathematical Sciences, Queen Mary University of London, UK\\
%	\vspace{-3mm}
%	$^3$ Department of Chemical Engineering and Biotechnology, University of Cambridge, UK}}

\maketitle

\abstract*{ Velocity-encoded MRI is an imaging technique used in different areas to assess flow motion. Some applications include medical imaging such as cardiovascular blood flow studies, and industrial settings in the areas of rheology, pipe flows, and reactor hydrodynamics, where the goal is to characterise dynamic components of some quantity of interest. 
The problem of estimating velocities from such measurements is a nonlinear dynamic inverse problem. To retrieve time-dependent velocity information, careful mathematical modelling and appropriate regularisation is required. 
In this work, we propose an optimisation algorithm based on non-convex Bregman iteration to jointly estimate velocity-, magnitude- and segmentation-information for the application of bubbly flow imaging. Furthermore, we demonstrate through numerical experiments on synthetic and real data that the joint model improves velocity, magnitude and segmentation over a classical sequential approach.}

\abstract{Velocity-encoded MRI is an imaging technique used in different areas to assess flow motion. Some applications include medical imaging such as cardiovascular blood flow studies, and industrial settings in the areas of rheology, pipe flows, and reactor hydrodynamics, where the goal is to characterise dynamic components of some quantity of interest. 
The problem of estimating velocities from such measurements is a nonlinear dynamic inverse problem. To retrieve time-dependent velocity information, careful mathematical modelling and appropriate regularisation is required. 
In this work, we propose an optimisation algorithm based on non-convex Bregman iteration to jointly estimate velocity-, magnitude- and segmentation-information for the application of bubbly flow imaging. Furthermore, we demonstrate through numerical experiments on synthetic and real data that the joint model improves velocity, magnitude and segmentation over a classical sequential approach.}
%This work deals with the problem of retrieving magnitude and phase images from dynamic and undersampled velocity-encoded MRI data. The typical approach to solve this problem is to separately estimate the magnitude image in a classical variational framework, for example using the edge-preserving total variation (TV) as regulariser, and subsequently derive the phase image from it. In this work, we propose a new approach to simultaneously reconstruct magnitude and phase images from time-dependent data. 
\section{Introduction}
\label{sec:1}
Magnetic resonance imaging (MRI) is an imaging technique that allows to visualise the chemical composition of patients or materials in a non-invasive fashion. Besides resolving in great detail the morphology of the object under consideration, MRI is intrinsically sensitive to motion, flow and diffusion  \cite{Burt19821044,Axel1984}.  %add more. 
This means that in a single experiment, MRI can produce both structural and functional information. % 
By designing the acquisition protocol appropriately, MRI can provide flow and motion estimation. This technique is known as MR velocimetry or phase-encoded MR velocity imaging \cite{Callaghan,doi:10.1146/annurev.fluid.31.1.95,Elkins2007,GLADDEN20132}. In this work, we will focus on the dynamic inverse problem involved in recovering velocities from this kind of data. 

%Phase-encoded MR velocity imaging is an imaging technique used in different areas to assess flow motion. 
In many MRI applications, the goal is not only to extract the structure of the object of interest, but also to estimate some functional features. An example is flow imaging, in which the aim is to reconstruct the velocity of the fluid that is moving in some structure. In order to acquire the velocity information and assess flow motion, phase-encoded MR velocity imaging is widely used in different areas. In medical imaging, this is used for example in cardiovascular blood flow studies to assess the distribution and variation in flow in blood vessels around the heart \cite{Gatehouse2005}. Other industrial applications include the study the rheology of complex fluids \cite{Callaghan1999}, liquids and gases flowing through packed beds \cite{Sederman1998,Sankey2009,Holland2010}, granular flows \cite{Fukushima99,Holland2008} and multiphase turbulence \cite{Tayler2012}. 

MRI scanners use strong magnetic fields and radio waves to excite subatomic particles (like protons) that subsequently emit radio frequency signals which can be measured by the radio frequency coils. Because the local magnetisation of the spins is a vector quantity, it is possible to derive both magnitude and phase images from the signal. Furthermore, for appropriately designed experiments, the velocity information can be estimated from the phase image. 
The problem of retrieving magnitude and phase (and therefore velocities) from such measurements is non-linear. Many standard approaches reduce this inverse problem to a complex but linear inverse problem, where magnitude and phase are estimated subsequently. With this strategy, however, it is impossible to impose regularity on the velocity information. In this work, we therefore propose a joint framework to simultaneously estimate magnitude and phase from undersampled velocity-encoded MRI.  %The underlying idea is impose regularity on  % to exploit the strong correlation between phase and magnitude and the time information, so that the reconstruction of both quantities will be improved. 
Based on \cite{Corona}, we additionally introduce a third task, that is the segmentation on the magnitude, to improve the overall reconstruction quality. The main motivation is that by estimating edges simultaneously from the data, both magnitude and segmentation are reconstructed more accurately. By enhancing the magnitude reconstruction, we expect in turn to improve the corresponding phase image and therefore the final velocity estimation.\\

\paragraph{Contributions} In this work we consider the problem of estimating flow, magnitude and segmentation of regions of interest from undersampled velocity-encoded  MRI data. The problem is of great interest in different areas including cardiovascular blood flow analysis in medical imaging and rheology of complex fluids in industrial applications.
To this end, we propose a joint variational model for undersampled velocity-encoded MRI. The significance of our approach is that by tackling the phase and magnitude reconstruction \textit{jointly}, we can exploit their strong correlation and finally impose regularity on the velocity component. This is further assisted by the introduction of a segmentation term as additional prior to enhance edges of the regions of interest. Our main contributions are
\begin{itemize}
\item A description of the forward and inverse problem of velocity-encoded MRI in the setting of bubbly flow estimation.
    \item A joint variational framework for the approximation of the non-linear inverse problem of velocity estimation. We show that by exploiting the strong correlation in the data, our joint method yields an accurate estimation of the underlying flow, alongside a magnitude reconstruction that 
preserves and enhances intrinsic structures and edges, due to a
joint segmentation approach. Moreover, we achieve an accurate segmentation to discern between different areas of interest, e.g. fluid and air. %In this unified Bregman iteration framework, we have the advantage of improving the reconstruction by reducing the contrast bias in the TV formulation, which leads to more accurate segmentation. In addition, the segmentation constitutes another prior for the reconstruction by enhancing edges of the regions of interest. Furthermore, we propose a non-convex alternating direction algorithm in a Bregman iteration scheme for which we prove global convergence.
\item An alternating Bregman iteration method for non-convex optimisation problems.
\item Numerical experiments on synthetic and real data in which we demonstrate the suitability and potential of our approach and provide a comparison with sequential approach.
\end{itemize}

\paragraph{Organisation of the paper} This paper is organised as follows. In Section~\ref{sec:velocityMRI} we describe the derivation of the inverse problem of velocity-encoded MRI from the acquisition process to the spin proton density estimation. In Section~\ref{sec:model} we present our joint variational model to jointly estimate phase and magnitude reconstruction and its segmentation. In Section~\ref{sec:optimisation} we propose an optimisation scheme to solve the non-convex and non-linear problem using Bregman iteration. To conclude, in Section~\ref{sec:results} we demonstrate the performance of our proposed joint method in comparison with a sequential approach for synthetic and real MRI data. 

\section{Velocity-encoded MRI}
\label{sec:velocityMRI}
In the following we will briefly describe the mathematics of the acquisition process involved in MRI velocimetry. Subsequently we are going to see that finding the unknown spin proton density basically leads to solving the inverse problem of the Fourier transform.
\subsection{From the Bloch equations to the inverse problem}
The magnetisation of a so-called spin isochromat can be described by the Bloch equations
\begin{align}
\frac{d}{dt}\left( \begin{array}{c} M_x(t)\\ M_y(t)\\ M_z(t)\end{array}\right) = \left( \begin{array}{ccc} -\frac{1}{T_2} & \gamma B_z(t) & -\gamma B_y(t)\\ -\gamma B_z(t) & -\frac{1}{T_2} & \gamma B_x(t)\\ \gamma B_y(t) & -\gamma B_x(t) & -\frac{1}{T_1} \end{array}\right)\left( \begin{array}{c}M_x(t)\\ M_y(t)\\ M_z(t) \end{array}\right) + \left( \begin{array}{c}0\\ 0\\ \frac{M_0}{T_1} \end{array}\right) \, \text{.}\label{eq:bloch1}
\end{align}
Here $M(t) = (M_x(t), M_y(t), M_z(t))$ is the nuclear magnetisation (of the spin isochromat), $\gamma$ is the gyromagnetic ratio, $B(t) = (B_x(t), B_y(t), B_z(t))$ denotes the magnetic field experienced by the nuclei, $T_1$ is the longitudinal and $T_2$ the transverse relaxation time and $M_0$ the magnetisation in thermal equilibrium. If we define $M_{xy}(t) = M_x(t) + i M_y(t)$ and $B_{xy}(t) = B_x(t) + i B_y(t)$, we can rewrite \eqref{eq:bloch1} to
\begin{subequations}
\begin{align}
\frac{d}{dt} M_{xy}(t) &= -i \gamma \left( M_{xy}(t) B_z(t) - M_z(t) B_{xy}(t) \right) - \frac{M_{xy}(t)}{T_2}\\
\frac{d}{dt} M_{z}(t) &= i \frac{\gamma}{2} \left( M_{xy}(t) \overline{B_{xy}}(t) - \overline{M_{xy}}(t) B_{xy}(t) \right) - \frac{M_{z}(t) - M_0}{T_1}
\end{align}\label{eq:complexbloch1}
\end{subequations}
with $\overline{\cdot}$ denoting the complex conjugate of $\cdot$.

If we assume for instance that $B = (0, 0, B_0)$ is just a constant magnetic field in $z$-direction, \eqref{eq:complexbloch1} reduces to the decoupled equations
\begin{subequations}
\begin{align}
\frac{d}{dt} M_{xy}(t) &= -i \gamma B_0 M_{xy}(t) - \frac{M_{xy}(t)}{T_2} \, \text{,}\label{subeq:comlexbloch21}\\
\frac{d}{dt} M_{z}(t) &= - \frac{M_{z}(t) - M_0}{T_1} \, \text{.}
\end{align}\label{eq:complexbloch2}
\end{subequations}
It is easy to see that this system of equations \eqref{eq:complexbloch2} has the unique solution
\begin{subequations}
\begin{align}
M_{xy}(t) &= e^{-t(i \omega_0 + 1/T_2)}M_{xy}(\Delta t)\label{eq:xymaghombckmag}\\
M_z(t) &= M_z(\Delta t) e^{-\frac{t}{T_1}} + M_0\left(1 - e^{-\frac{t}{T_1}}\right)
\end{align}
\end{subequations}
for $\omega_0 := \gamma B_0$ denoting the Lamor frequency, and $M_{xy}(\Delta t)$, $M_z(\Delta t)$ being the initial magnetisations at time $t = \Delta t$.

\subsection{Signal recovery}\label{subsec:signalrecovery}

The key idea to enable spatially resolved nuclear magnetic resonance spectrometry is to add a magnetic field $\hat{B}(t)$ to the constant magnetic field $B_0$ in $z$-direction that varies spatially over time. Then, \eqref{subeq:comlexbloch21} changes to
\begin{align}
\frac{d}{dt} M_{xy}(t) &= -i\gamma(B_0 + \hat{B}(t))M_{xy}(t) - \frac{M_{xy}(t)}{T_2} \, \text{,}\nonumber
\intertext{which, for initial value $M_{xy}(\Delta t)$, has the unique solution}
M_{xy}(t) &= e^{-i \gamma \left(B_0 t + \int_{\Delta t}^t \hat{B}(\tau) \, d\tau\right)} e^{-\frac{t}{T_2}} M_{xy}(\Delta t) \, ,
\end{align}
if we ensure $\hat{B}(\Delta t) = 0$. If now $x(t)$ denotes the spatial location of a considered spin isochromat at time $t$, we can write $\hat{B}(t)$ as $\hat{B}(t) = x(t) \cdot g(t)$, with a function $g$ that describes the influence of the magnetic field gradient over time. 

If a radio-frequency (RF) pulse that has been used to induce magnetisation in the $x$-$y$-plane is subsequently turned off at time $t = t_\ast$ and thus, $B_x(t) = 0$ and $B_y(t) = 0$ for $t > t_\ast > \Delta t$, the same coils that have been used to induce the RF pulse can be used to measure the $x$-$y$ magnetisation. Using \eqref{eq:xymaghombckmag} and assuming $t_\ast < t \ll T_2$ for all $x \in \mathbb{R}^3$, this gives rise to the following model-equation:
\begin{align}
M_{xy}(t) &= e^{-i \gamma \left(B_0 t + \int_{\Delta t}^t x(\tau) \cdot g(\tau)  \, d\tau\right)} M_{xy}({\Delta t}) \, .\label{eq:nmrsigacq}
\end{align}

In the following we assume that $x(t)$ can be approximated reasonably well via its Taylor approximation around $t = \Delta t$, i.e.
\begin{align*}
    x(t) = \sum_{n = 0}^\infty \frac{x^{(n)}(\Delta t)}{n !} t^n \, ,
\end{align*}
which yields
\begin{align}
\int_{\Delta t}^t x(\tau) \cdot g(\tau) \, d\tau = \sum_{n = 0}^\infty \left[\frac{x^{(n)}({\Delta t})}{n !} \cdot \int_{\Delta t}^t g(\tau) \, \tau^n \, d\tau \right] \, \text{.}\label{eq:taylapprox}
\end{align} 
It is well-known that appropriate application of gradients (i.e. appropriate design of $g$) enables the approximation of individual moments of \eqref{eq:taylapprox}. If we further assume that the system to be observed does only contain zero- and first-order moments, we can assume 
\begin{align}
\int_{\Delta t}^t x(\tau) \cdot g(\tau) \, d\tau = x \cdot \int_{\Delta t}^t g(\tau) \, d\tau + \varphi \cdot \int_{\Delta t}^t g(\tau) \tau \, d\tau\, \text{,}
\label{eq:taylapprox1}
\end{align} 
where $x$ is now short for $x({\Delta t})$ and $\varphi := x^\prime ({\Delta t})$ is the corresponding velocity information.\\

In order to turn \eqref{eq:nmrsigacq} into a useful mathematical model we need to encode velocity information and remove the temporal dependency of $x$. In order to do so, the gradient coils need to be programmed to first enable the encoding of velocity information via a function $g$ that satisfies
\begin{align*}
    \int_{\Delta t}^{t} g(\tau) \, d\tau = 0 \qquad \text{and} \qquad \int_{\Delta t}^{t} g(\tau) \tau \, d\tau = \zeta(t) \, ,
\end{align*}
for time $\Delta t \leq t \leq t_1$ and a function $\zeta:\mathbb{R}_{\geq 0} \mapsto \mathbb{R}^3$. Subsequently, the gradients need to be programmed to encode spatial information by choosing $g$ with
\begin{align*}
    \int_{t_2}^{t} g(\tau) \, d\tau = \xi(t) \qquad \text{and} \qquad \int_{t_2}^{t} g(\tau) \tau \, d\tau = 0 \, ,
\end{align*}
for $t_2 > t_1$ and a function $\xi : \mathbb{R}_{\geq 0} \mapsto \mathbb{R}^3$. Since the RF-coils measure a volume of the whole $x$-$y$ net-magnetisation, the acquired signal then equals
\begin{align}
    f(t) = \int_{\mathbb{R}^3} u(x) \, e^{-i \gamma \left(B_0(x) t + \varphi(x) \cdot \zeta(t) \right)} \, e^{- i \gamma x \cdot \xi(t)} \, dx \, .\label{eq:simpleforwardmodel}
\end{align}
with $u(x)$ denoting the spin-proton density $M_{xy}({\Delta t})$ at a specific spatial coordinate $x \in \mathbb{R}^3$. Note that for $r(x) := u(x) \, e^{-i \gamma \left(B_0(x) t + \varphi(x) \cdot \zeta(t) \right)}$ we observe that $f$ is just the Fourier transform of the complex signal $r$ with magnitude $u$ and phase $-\gamma ( B_0 t + \varphi \cdot \zeta )$.

\subsection{Removal of background magnetic field}
\label{sec:flowphase}
Our goal is to recover the velocity information $\varphi$ from $f$. Assuming that we do not know $B_0$, we can alternatively conduct two experiments, where the setup is identical apart from the velocity-encoding gradients having opposite polarities, i.e. we measure
\begin{subequations}
\begin{align}
    f_{+}(t) &= \int_{\mathbb{R}^3} u(x) \, e^{-i \gamma \left(B_0(x) t + \varphi(x) \cdot \zeta(t) \right)} \, e^{- i \gamma x \cdot \xi(t)} \, dx \, ,\\
    f_{-}(t) &= \int_{\mathbb{R}^3} u(x) \, e^{-i \gamma \left(B_0(x) t - \varphi(x) \cdot \zeta(t) \right)} \, e^{- i \gamma x \cdot \xi(t)} \, dx \, .
\end{align}\label{eq:two_measurements}
\end{subequations}
Hence, if we denote $\varphi_{+}(x, t) := B_0(x) t + \varphi(x) \cdot \zeta(t)$ and $\varphi_{-}(x, t) := B_0(x) t - \varphi(x) \cdot \zeta(t)$, we immediately observe 
\begin{align*}
    \varphi(x) \cdot \zeta(t) = \frac12 \left( \varphi_{+}(x, t) - \varphi_{-}(x, t) \right) \, .
\end{align*}
The inverse problem of \eqref{eq:two_measurements} is to recover $u$ and $\varphi$ from $f_+$ and $f_-$.
% \begin{figure}[t]
% \centering
% \includegraphics[width=0.7\textwidth]{2vel.png}
% \caption{Illustration of the acquisition of two phase images. We can see that $g_z(t)$ have opposite polarities.}
% \label{fig:2vel}
% \end{figure}
\subsection{Zero-flow experiment}
\label{sec:zeroflow}
%\noindent \textcolor{red}{This section requires more accurate information on why we require these additional measurements.}
%From \eqref{eq:taylapprox1} we can see that the velocity $\varphi$ is proportional to the amplitude and duration of the gradient and therefore can be controlled by varying these two settings in the acquisition. However, in practice, background phase effects such as field inhomogeneity, sampling offsets and inaccurate timing of the gradients can perturb the phase and cause measurement errors.\\
A zero-flow experiment that allows for the removal of additional artefacts is also conducted. This experiment is to account for imperfections in the measurement system which cause an added signal between the positive and negative $\zeta$ experiments even in the absence of flow, and enables a correction that allows direct quantification of flow and tissue motion.
%A zero-flow experiment allows to remove additional artifacts that stem from imperfections of the measurement system, background effects and allows direct quantification of flow and tissue motion. 
We refer to this technique as flow compensation, %\textcolor{red}{Is this really the main reason?} 
which consists of acquiring a reference scan, with any flow switched-off, with vanishing zero and first gradient moments, before the actual velocity encoding scan with added bipolar gradients is performed. In this way, we obtain background phase images from the reference scan, and velocity sensitivity with the second flow-sensitive scan. 
In practice, this means that in addition to \eqref{eq:two_measurements}, the following two measurements are taken: 
%\textcolor{blue}{Shouldnt it be something like noflow instead of flow? Flow compensation is not actually estimating the velocity but rather the background effects }
\begin{subequations}
\begin{align}
    f_{\text{noflow}_+}(t) &= \int_{\mathbb{R}^3} u(x) \, e^{-i \gamma \varphi_{\text{noflow}_+}(x, t)} \, e^{- i \gamma x \cdot \xi(t)} \, dx \, ,\\
    f_{\text{noflow}_-}(t) &= \int_{\mathbb{R}^3} u(x) \, e^{-i \gamma \varphi_{\text{noflow}_{-}}(x, t)} \, e^{- i \gamma x \cdot \xi(t)} \, dx \, ,
\end{align}\label{eq:four_experiments}
\end{subequations}
so that the actual velocity information can be recovered via
\begin{align}
    \varphi(x) \cdot \zeta(t) = \frac12 \left( \left( \varphi_{+}(x, t) - \varphi_{-}(x, t) \right) - \left( \varphi_{\text{noflow}_{+}}(x, t) - \varphi_{\text{noflow}_{-}}(x, t) \right) \right) \, .\label{eq:velocity_equation}
\end{align}
The inverse problem is to recover $u$ and $\varphi$ from \eqref{eq:two_measurements} and \eqref{eq:four_experiments} via \eqref{eq:velocity_equation}. More details on phase-encoded MR velocity imaging can be found in \cite{Markl20061VE}.\\

%\noindent\textcolor{red}{Veronica, can you merge the first part of this subsection with the remainder of this section? Thank you!}

% \noindent \textcolor{red}{I think we need to make it clear that the signal acquisition is two-fold: first the velocity-encoding (from time $[0, t_1]$?), which will encode the phase. Then the spatial-encoding (from $[t_2, t]$ with $t_2 > t_1$?), which will ultimately lead to an equation similar to the following:
% \begin{align*}
%     y(t) = \int_{\mathbb{R}^3} u(x) e^{-i \gamma \left(B_0(x) t + \varphi(x) \cdot \int_0^{t_1} g(\tau) \, \tau \, d\tau \right)} e^{- i x \cdot \xi(t)} \, dx \, ,
% \end{align*}
% for $\xi(t) := \gamma \int_{t_2}^t g(\tau) \, d\tau$.}
%\newline
In other words, for a given direction of the velocity to be measured ($x$, $y$ or $z$), the corresponding component velocity map ($v_x$, $v_y$ or $v_z$) is acquired by applying repeatedly a pulse sequence with the velocity-encoding gradient in the respective direction ($x$, $y$ or $z$) and with alternating polarity between consecutive pulse sequences (from $\pm g$ to $\mp g$). The difference between the phase of the MRI image reconstructed from the acquired k-space data of consecutive pulse sequences, and the reference to a zero flow experiment, yields the component velocity map.

\subsection{Sampling}
The MRI signal is acquired by sampling the continuous signals of $f_+$, $f_-$, $f_{\text{noflow}_+}$ and $f_{\text{noflow}_-}$ at $m$ discrete points in time. Hence, for each phase $-\phi$ the data acquisition reads as
\begin{align}
    f_j = \int_0^{t_\ast} \Psi(t, t_j) \left[ \int_{\mathbb{R}^3} u(x) \, e^{i \, \phi(x, t)} \, e^{- i \gamma x \cdot \xi(t)} \, dx \right] \, dt \, ,\label{eq:sampling1}
\end{align}
for $j \in \{1, \ldots, m\}$ and where $\Psi$ denotes the sampling function or distribution. If we for example assume $\Psi(t, t_j) = \delta(t - t_j)$, where $\delta$ denotes the Dirac delta distribution, and that $\phi(x, t)$ is constant w.r.t. time, then \eqref{eq:sampling1} simplifies to
\begin{align}
    f_j = \int_{\mathbb{R}^3} u(x) \, e^{i \, \phi(x)} \, e^{-i \gamma x \cdot \xi_j} \, dx  \, , \qquad j \in \{1, \ldots, m\} \, . \label{eq:sampling2}
\end{align}
We want to denote the sub-sampled Fourier transform with $SF$, and therefore rewrite \eqref{eq:sampling2} to 
\begin{align}
    f = SF \left( u e^{i \phi} \right) \, ,\label{eq:forward_problem}
\end{align}
where $f \in \mathbb{R}^m$ denotes the vector of k-space samples. 
Sampling strategies are very important to reduce the acquisition times and therefore to be able to image dynamic systems using velocity-encoded MRI through fast imaging techniques. The main idea is to exploit redundancy in some specific domain of the measured data. This approach is strongly related to the theory of compressed sensing (CS) \cite{Candes2004,Donoho2006,Lustig2007} and many image reconstruction techniques have been proposed \cite{Holland2010,Tayler2012,Jung2008,Paciok2011,Paulsen2010,Parasoglou2009}. \\
Depending on whether $\gamma \xi$ is sampled on a uniform or non-uniform grid, $SF$ can be realised via the Fast Fourier Transform (FFT) \cite{fft} or via a non-uniform Fourier Transform such as NUFFT \cite{NUFFT}.

\subsection{Dynamic inverse problem}
We want to highlight that every $u$ and $\phi$ in \eqref{eq:forward_problem} implicitly depends on an initial time $\Delta t$, which becomes evident from the derivation in Section \ref{subsec:signalrecovery}. Hence, if we take measurements for a sequence $\{ \Delta t_j \}_{j = 1}^s$ with $0 = \Delta t_1 < \Delta t_2 < \ldots < \Delta t_s$, we are introducing a discrete temporal dimension to our inverse problem that potentially allows us to exploit any temporal correlation between frames $\{ u_j \}_{j = 1}^s$ and $\{ \phi_j \}_{j = 1}^s$. However, we will only consider the reconstruction of individual frames throughout this work for reasons that we are going to address later.

% stored as a matrix in the so-called k-space, computationally we can exploit the Fast Fourier Transform (FFT) method \cite{fft}. %In the discrete setting and under the presence of noise, the k-space matrix Y, vectorised as Yv, is given by: where is the discrete Fourier transform operator, Uv is the vectorised image and Ev is a vector of noise elements. 
In the following we will refer to an individual frame of the dynamic inverse problem for velocity-encoded MRI in the discrete setting and under the presence of noise making use of the notation of the discrete Fourier transform operator. %In particular, $u, \phi \in \mathbb{R}^n$ are the proton density or magnitude image and correspondent phase image, respectively, in an image domain $\Omega:=\{1,\dots,n_1 \} \times \{1,\dots,n_2 \}$, with $n=n_1n_2$. The vector $y=(y_i)_{i=1}^m\in \mathbb{C}^m$  with $m<<n$ are the measured Fourier coefficients obtained from \eqref{eq:forward_problem}.

%\noindent \textcolor{red}{Should we say anything about the discretisation of $u$ and $\phi$}
%\textcolor{blue}{Is it enough what we write next or do you think we need more? }

\section{Mathematical modelling}
\label{sec:model}
In this section we first present the velocity-encoded MRI reconstruction inverse problem in the presence of noise and discuss a sequential variational regularisation scheme to approximate the solution. Secondly, we introduce our joint reconstruction and segmentation approach in a Bregman iteration framework to jointly estimate phase, magnitude and segmentation.
\subsection{Indirect phase-encoded MR velocity imaging}
% To be able to image dynamic systems using velocity-encoded MRI, fast imaging techniques are required to reduce the acquisition times. One way to achieve this is by exploiting redundancy of the measured data, thus performing undersampling. This approach is strongly related to the theory of compressed sensing (CS) \cite{Candes2004,Donoho2006,Lustig2007} and many image reconstruction techniques have been proposed \cite{Holland2010,Tayler2012,Jung2008,Paciok2011,Paulsen2010,Parasoglou2009}. 
The velocity-encoded MRI image reconstruction problem is described as follows. % Let $u$ be the magnitude image and $\phi$ the phase image for an individual measurement $y$.
Let $u, \phi \in \mathbb{R}^n$ be the proton density or magnitude image and correspondent phase image, respectively, in a discretised image domain $\Omega:=\{1,\dots,n_1 \} \times \{1,\dots,n_2 \}$, with $n=n_1n_2$. The vector $f=(f_i)_{i=1}^m\in \mathbb{C}^m$  with $m \ll n$ are the measured Fourier coefficients obtained from \eqref{eq:forward_problem}.
Based on \eqref{eq:forward_problem} the forward model for noisy data is given by
\begin{equation}
f=SF\left(u e^{i \phi}\right) + \eta \, ,
\label{eq:forward}
\end{equation}
where % $\cdot$ is the element-wise multiplication operator, 
$i^2=-1$ and $\eta$ is Gaussian noise with zero mean and standard deviation $\sigma$. For brevity we will follow the notation $A=SF$. As explained in the previous section, velocity information is encoded in the phase image. However, during the acquisition the phase is perturbed by an error due to field inhomogeneity and chemical shift. To account for this error, usually different measurements corresponding to different polarities of encoding flow  gradients are acquired. Then the velocity (in one direction) at one particular time will be estimated as in \eqref{eq:velocity_equation}, where $\zeta$ is a constant known from the acquisition setting.
% \begin{equation}
%  \varphi(x) = \frac{\phi_1(x) - \phi_2(x)}{\text{const}} , 
%  \label{eq:phasediff}
%  \end{equation}
 %in the case of two measurements, where $const$ is known from the acquisition sequence settings. 

Given the presence of noise and partial observation of the data due to undersampling, the problem described in \eqref{eq:forward} is ill-posed. A simple strategy to obtain an approximated solution is to replace with zero the missing Fourier coefficients and compute the so-called zero-filling solution
\begin{equation}
r_z = A^*f
\end{equation}
where $r=u e^{i \phi}$. However, these reconstructed images will present aliasing artefacts because of the undersampling. A classical approach to solve this problem is to compute approximate solutions of \eqref{eq:forward} using a variational regularisation approach. We consider a Tikhonov-type regularisation approach that reads
\begin{equation}
r_j \in \argmin{r} \Big\{ \frac{1}{2} \| A_j r - f_j\|_{2}^{2} + \alpha  J(r) \Big\}, 
\label{eq:tik}
\end{equation}
for $j\in\{1,\dots, 4\}$ being the different measurements, where the first term is the data fidelity that imposes consistency between the reconstruction and the given measurements $f$, the second term is the regularisation, which incorporates some prior knowledge of the solution. The parameter $\alpha > 0$ is a regularisation parameter that balances the two terms in the variational scheme. %In this setting, different regularisation functionals $J$ can be chosen (see \cite{BenningGladdenHollandEtAl2014} for a survey of variational regularisation approaches). 
In this setting, the survey proposed in \cite{BenningGladdenHollandEtAl2014} describes different choices for the regularisation functional $J$, including wavelets and higher-order total variation (TV) schemes.
Subsequently, the phases can be extracted from these complex images $r_j=u_j e^{i \phi_{j}}$ as
\begin{equation}
 \phi_j=\arg(r_j).
\end{equation}

More recently, other reconstruction approaches have been proposed to regularise the phase of the image \cite{Fessler2004,Zibetti2010,Zhao2012,valkonen2014primal,Zibetti:2017}. All these methods rely on modelling separately prior knowledge on the magnitude and on phase images and differ on the optimisation schemes involved in the non-convex and non-linear problem. However, while it is possible to exploit information about the velocity from fluid mechanics, it is in general hard to assume specific knowledge on the individual phases. As explained in the previous section and described in \eqref{eq:velocity_equation}, velocities are computed as phase differences of different MR measurements and therefore the regularisation needs to be imposed on the phase difference rather than individual phases. In this work, we step away from the approach of only regularising individual phases and propose instead to regularise the velocity as difference of phases. % \textcolor{red}{We should probably remove the last sentence as we don't penalise the difference of phases but the phases individually}. 
In the following we describe our choice of regularisation and algorithmic framework for velocity-encoded MRI. 
%This work deals with the problem of retrieving magnitude and phase images from dynamic and undersampled velocity-encoded MRI data. The typical approach to solve this problem is to separately estimate the magnitude image in a classical variational framework, for example using the edge-preserving total variation (TV) as regulariser, and subsequently derive the phase image from it. In this work, we propose a new approach to simultaneously reconstruct magnitude and phase images from time-dependent data. 

\subsection{Joint variational model}
In many industrial applications, velocity-encoded MRI is used to estimate flow of different chemical species in different physical status, such as gas-liquid systems \cite{Gladden2017}. In this case, one aims at recovering a piecewise constant image or an image with sharp edges to facilitate further analysis such as identification of regions of interest. 
It was proposed in \cite{Corona} to use a segmentation task as additional regularisation on the reconstruction to impose regularity in terms of sharp edges. It was shown there that this is highly beneficial for very low undersampling rates in MRI. In this work, we expand this idea to the phase-encoded MR velocity imaging data, where the idea is to jointly solve for magnitude, segmentation and phase improving performances on the three tasks. 

%As discussed in the previous section, to acquire a one component velocity image, four MR images are needed, corresponding to: flow and $\pm g$ polarity for the velocity-encoding gradient ($u_1 e^{i \phi_1}$); flow and $\mp g$ polarity the velocity-encoding gradient ($u_2 e^{i \phi_2}$); no flow and $\pm g$ polarity for the velocity-encoding gradient and ($u_3 e^{i \phi_3}$); and no flow and $\mp g$ polarity the velocity-encoding gradient ($u_4 e^{i \phi_4}$).

Following the work in \cite{Corona}, we are interested in the \textit{joint model} to recover magnitude $u_j$ and velocity $\varphi$ components through the measured phases $\phi_j$ from undersampled MRI data $f_i$ and to estimate a segmentation $v_j$ on the magnitude images. As described in the previous section, we are dealing with four MRI measurements to obtain one component velocity image. Defining the shorthand notations $u := \{u_j\}_{j = 1}^4$, $v := \{ v_j \}_{j = 1}^4$ and $\phi := \{ \phi_{j} \}_{j = 1}^4$, this joint model reads as
\begin{equation}
\begin{aligned}
E(u, v, \phi) {} = {} &\sum_{j=1}^4 \Bigg\{ \frac{1}{2} \underbrace{\| A(u_j e^{i \phi_i}) - f_j \|_2^2}_{\text{reconstruction}} \\
&+ \delta \underbrace{\sum_n v_{nj}(c_1 - u_{nj})^2 + (1-v_{nj})(c_2 - u_{nj})^2 \Bigg\} }_{\text{segmentation}} \, .
\end{aligned}
\label{eq:joint}
\end{equation}
\noindent %\textcolor{red}{$k$ as an index over the spatial coordinates is also sub-optimal as $k$ later denotes the iteration index.}
%with $u=(u_1, \dots,u_4)$, $v=(v_1, \dots,v_4)$, $\varphi=(\varphi_1,, \dots,\varphi_4)$ and $f=(f_1, \dots,f_4)$. % and $c_1,c_2$ are the constants for the segmentation. %D_J^p(x,\tilde{x})=J(x)-J(\tilde{x}) - \langle p, x - \tilde{x}\rangle$ is the Bregman distance for the regularisation terms $J_1=\| \nabla u \|$, $J_2=\| \nabla v \|$ and $J_3=\| \nabla(\varphi_1 - \varphi_2)  \|$.
%The joint cost function is non convex. The sub-problems in $u$ and $v$ are convex, but the sub-problem in $\varphi$ is non linear and non convex and we solve it following \cite{LBREG,Benning2017} using a linearised Bregman iteration. 
%For $k=1,\dots, maxIter$, we perform alternating minimisation in each variable 
%\begin{equation}
%\begin{aligned}
%u^{k+1} &= \text{arg}\min_u \Big\{ \frac{1}{2} \| A(u e^{i \varphi^{k}}) - f \|_2^2 + \alpha D_{J_1}^{p^k} (u,u^k)\\
%&\qquad \qquad ~ ~ +\delta \sum_\Omega v^{k}(c_1 - u)^2 + (1-v^{k})(c_2 - u)^2    \Big\}\\
%p^{k+1} &= p^{k} -\frac{1}{\alpha} (A^* (A(u^{k+1} e^{i \varphi^{k}}) - f) + 2\delta(v^{k}(u^{k+1}-c_1) \\
%& + (1-v^{k})(u^{k+1}-c_2) ))\\
%v^{k+1} &= \text{arg}\min_v \Big\{ v\big((c_1 - u^{k+1})^2 -(c_2 - u^{k+1})^2\big) + \beta D_{J_2}^{q^k}(v,v^k) \Big\}\\
%q^{k+1} &= q^{k} -\frac{1}{\beta} ((c_1 - u^{k+1})^2 -(c_2 - u^{k+1})^2 ) \\
%\varphi^{k+1}  &=  \text{arg}\min_{\varphi} \Big\{\tau \langle \varphi -\varphi^k, \nabla (A(u e^{i \varphi}) - f)\rangle + \gamma D_{J_3}^{r^k} (\varphi, \varphi^k) \Big\} \\
%r^{k+1} &= r^{k} - \frac{\tau}{\gamma} \nabla (A(u^{k+1} e^{i \varphi^k})- f)
%\end{aligned}
%\label{eq:joint}
%\end{equation}
The first term in \eqref{eq:joint} describes the reconstruction fidelity term for the magnitudes $u$ and phases $\phi$ for the given data $f := \{ f_j \}_{j = 1}^4$. The second term represents the segmentation problem to find partitions $v$ of the images $u$ in two disjoint regions that have mean intensity values close to the constants $c_1$ and $c_2$ \cite{Chant2001,Chan2006}. The parameter $\delta$ weighs the effect of the segmentation onto the reconstruction. 
The underlying idea is to exploit structure and redundancy in the data, estimating edges simultaneously from the data, ultimately improving the reconstruction. By incorporating prior knowledge of the regions of interest we impose additional regularity of the solution. 

The joint cost function  \eqref{eq:joint} is non-convex. While sub-problems in $u$ and $v$ (leaving the other parameters fixed) are convex, the sub-problems in $\phi$ are non-linear and non-convex. In the next section we present a unified framework based on non-convex Bregman iterations to solve the joint model.

\section{Optimisation}
\label{sec:optimisation}
There are many ways of minimising \eqref{eq:joint}. We want to pursue a strategy that guarantees smooth velocity-components, piecewise-constant segmentations and magnitude images with sharp transitions in an inverse scale-space fashion. In order to achieve those features, we aim to approximate minimisers of \eqref{eq:joint} via an alternating Bregman proximal method or Bregman iteration of the form
\begin{subequations}
\begin{align}
%\begin{split}
    u_l^{k + 1} &\in \arg\min_{u} \left\{ E(u_1^{k + 1}, \ldots, u_{l - 1}^{k + 1}, u, u_{l + 1}^k, \ldots, u_d^k, v^k, \phi^k) + D_{J_u}^{p_l^k}(u, u_l^k) \right\} \, , \\
    p_l^{k + 1} &= p_l^k - \frac{\partial}{\partial u_l} E(u_1^{k + 1}, \ldots, u_{l - 1}^{k + 1}, u_l^{k + 1}, u_{l + 1}^k, \ldots, u_d^k, v^k, \phi^k) \, , \\
    v_l^{k + 1} &= \argmin{v} \left\{ E(u^{k + 1}, v_1^{k + 1}, \ldots, v_{l - 1}^{k + 1}, v, v_{l + 1}^k, \ldots, v_d^k, \phi^k) + D_{J_v}^{q_l^k}(v, v^k_l) \right\} \, ,\\
    q_l^{k + 1} &= q^k_l - \frac{\partial}{\partial v_l} E(u^{k + 1}, v_1^{k + 1}, \ldots, v_{l - 1}^{k + 1}, v_l^{k + 1}, v_{l + 1}^k, \ldots, v_d^k, \phi^k) \, ,\\
    \phi^{k + 1} &= \argmin{\varphi} \left\{ \langle \partial_{\phi}  E(u^{k + 1}, v^{k + 1}, \phi^k), \phi \rangle + D_{J_{\phi}}^{w^k}(\phi, \phi^k) \right\} \, ,\\
    w^{k + 1} &= w^k - \frac{\partial}{\partial \phi}E(u^{k + 1}, v^{k + 1}, \phi^k) \, .
%\end{split}    
\end{align}\label{eq:bregmanproximal}
\end{subequations}
for $l = 1, \ldots, d := 4$, $u := (u_l)_{l = 1}^d$, $v := (v_l)_{l = 1}^d$ and $\phi := (\phi_l)_{l = 1}^d$. Here $J_u$, $J_v$ and $J_{\phi}$ are proper, lower semi-continuous and convex functions and $ D_{J_u}^{p_l^k}(u, u_l^k)$, $ D_{J_v}^{q_l^k}(v, v_l^k)$ and $ D_{J_{\phi}}^{w^k}(\phi, \phi^k)$ are the corresponding generalised Bregman distances \cite{bregman1967relaxation,kiwiel1997proximal} with arguments and corresponding subgradients $p_l^k$, $q_l^k$ and $w^k$. A generalised Bregman distance is the distance between a function $J$ evaluated at argument $u$ and its linearisation around argument $v$, i.e.
\begin{align*}
    D_J^q(u, v) = J(u) - J(v) - \langle q, u - v \rangle \, ,
\end{align*} 
for a subgradient $q \in \partial J(v)$. Note that algorithm \eqref{eq:bregmanproximal} has update rules for the subgradients, as $J_u$, $J_v$ and $J_{\phi}$ are allowed to be non-smooth, which makes the selection of particular subgradients necessary.

The algorithm is a hybrid of the algorithms proposed in \cite{Benning2017} and \cite{Corona}. For both algorithms global convergence results, motivated by \cite{xu2013block,bolte2014proximal}, have been established. %\textcolor{blue}{Do we need to say more explicitly that the convergence analysis hold for this hybrid case too?} 
Since we deal with imperfect data potentially corrupted by measurement noise and numerical errors, we will, however, use \eqref{eq:bregmanproximal} in combination with an early-stopping criterion in order not to converge to a minimiser of \eqref{eq:joint} but to approximate the solution of \eqref{eq:forward} via iterative regularisation. 

The crucial part for the application of \eqref{eq:bregmanproximal} are the choices of the underlying functions $J_u$, $J_v$ and $J_{\phi}$ of the corresponding Bregman distances. We want both the magnitude images and the segmentations to maintain sharp discontinuities and therefore want to penalise their discretised, isotropic, total variation. On the other hand, we want to guarantee smooth components of our velocity field, which is why we penalise them with the two-norm of a discretised gradient. In particular, we choose
\begin{align}
    J_u(u) = \alpha \, \text{TV}(u) := \alpha \| | \nabla u | \|_1 \, ,\qquad J_v(v) := \beta \, \text{TV}(v) ,\label{eq:totalvariation}
\end{align}
to be the isotropic total variation with weights $\alpha > 0$ and $\beta > 0$, where $\nabla :\R^n \rightarrow \R^{2n}$ denotes a forward finite-difference approximation of the gradient, $| \cdot |$ the Euclidean vector norm and $\| \cdot \|_1$ the pixel-wise one-norm. Further, we choose $J_{\phi}$ in a way that allows to enable an $H^1$-norm-type smoothing on the difference of the phases, i.e.
\begin{align*}
    J_{\phi}^{k + 1}(\phi) = \frac{1}{2 \tau} \left( \eta \| | \nabla (\phi_1 - \phi_2) - (\phi_3 - \phi_4 ) | \|^2 + \sum_{l = 1}^d \| \phi_l \|^2 \right)  \, ,
\end{align*}
where $\eta > 0$ denotes another weight. Note that all convex sub-optimisation-problems in \eqref{eq:bregmanproximal} are solved numerically with a primal-dual hybrid gradient (PDHG) method \cite{zhu2008efficient,pock2009algorithm,esser2010general,chambolle2011first}. Once we have approximated the magnitudes, labels and phases with this iterative regularisation strategy, we can compute the velocity components via \eqref{eq:velocity_equation}.
% To obtain a smooth velocity component from the phase differences, we penalise $u_{13}$ with a weighted $H^1$-norm, i.e.
% \begin{align*}
%     J_{13}(u) = \frac{1}{2} \left( \| u \|^2 + \alpha_i \| | \nabla u | \|^2 \right) \, .
% \end{align*}

% \noindent\textcolor{red}{Now I have a problem with $i$'s as indices ;-) I will change it later. Is this actually correct? Do you solve for $\varphi$ or does this happen subsequently? And do you smooth only $\varphi$, or also $\phi$?}
% \textcolor{blue}{oh yes this $i$ is annoying! I solve for $\varphi$ by combining the $\phi_i$s in a diagonal matrix in the code and only smooth $\varphi$ with the (non-squared!) 2-norm of the second order differences.}

\section{Numerical results}
\label{sec:results}
In this section we present numerical results of our method for the specific application of bubble burst hydrodynamics using MR velocimetry. The hydrodynamics of bursting bubbles is important in many different areas such as  geophysical processes and bioreactor design. We refer to \cite{Reci} for an overview on the field and the description of results on the first experimental measurement of the liquid velocity field map during the burst of a bubble at the liquid surface interface. %and used to investigate a discrepancy between the predictions of different numerical methods. 

%In this paper we use data acquired using MR velocimetry where the three velocity components are obtained on longitudinal and transverse views.

%In velocity-encoded MRI, undersampling is crucial to achieve fast imaging and being able to track the underlying dynamic system. Often the acquired k-space data is sampled on non-Cartesian grid to accelerate the coverage of the k-space. An example is spiral imaging and this requires a the use of the NUFFT operator. The experiments on real data are implemented using the NUFFT operator from \cite{NUFFT}. 

% \subsection{Sequential approach}
% \textcolor{blue}{This needs to be removed/integrated with Martin's part.}
% We will compare the performance of our proposed joint approach to a sequential approach where the magnitude is obtained by solving the variational problem
% \begin{equation}
% u_j = \argmin{u}  \frac{1}{2} \| A_j u - f_j\|_{2}^{2} + \alpha  \operatorname{TV}(u) , 
% \end{equation}
% for $j=\{1,\dots,4\}$ and sequentially computing the phase images using the linearised Bregman iteration as in \cite{LBREG,Benning2017}
% \begin{eqnarray}
% \varphi_j^{k+1} &=& \argmin{u} \Big\{ \tau \langle \varphi_j - \varphi_j^k, \nabla H(\varphi_j^k)\rangle + D_J^{p^k}(\varphi_j,\varphi_j^k) \Big\}\\
% p_j^{k+1}&=& p_j^k-\tau \nabla H(\varphi_k^k).
% \end{eqnarray}
\subsection{Case-study on simulated dataset}
To quantitatively evaluate our method, we consider the simulated k-space data of a rising spherical bubble in an infinite fluid during Stoke's flow regime. The simulated data consists of 32 time frames, but for the sake of compactness we will show some visual outputs for one time step $t=19$. %See \ref{app} for the full time resolution. 

We assess the performance of our approach for velocity and magnitude estimation by comparing our solutions with respect to the groundtruth and using the mean squared error (MSE) defined as $\| x^{\text{groundtruth}} - x\|_2^2 /n  $, where $n$ is the number of pixels in the image. 

We also present a comparison with a sequential approach, where the magnitude is obtained with a classic CS TV-regularised approach and the phase is subsequently estimated using the method proposed in \cite{Benning2017} and presented in \cite{Reci} for the evaluation of bubbly flow estimation. 

\begin{figure}[t]
   \centering
\subfloat[ Groundtruth]{\includegraphics[width=0.2\textwidth]{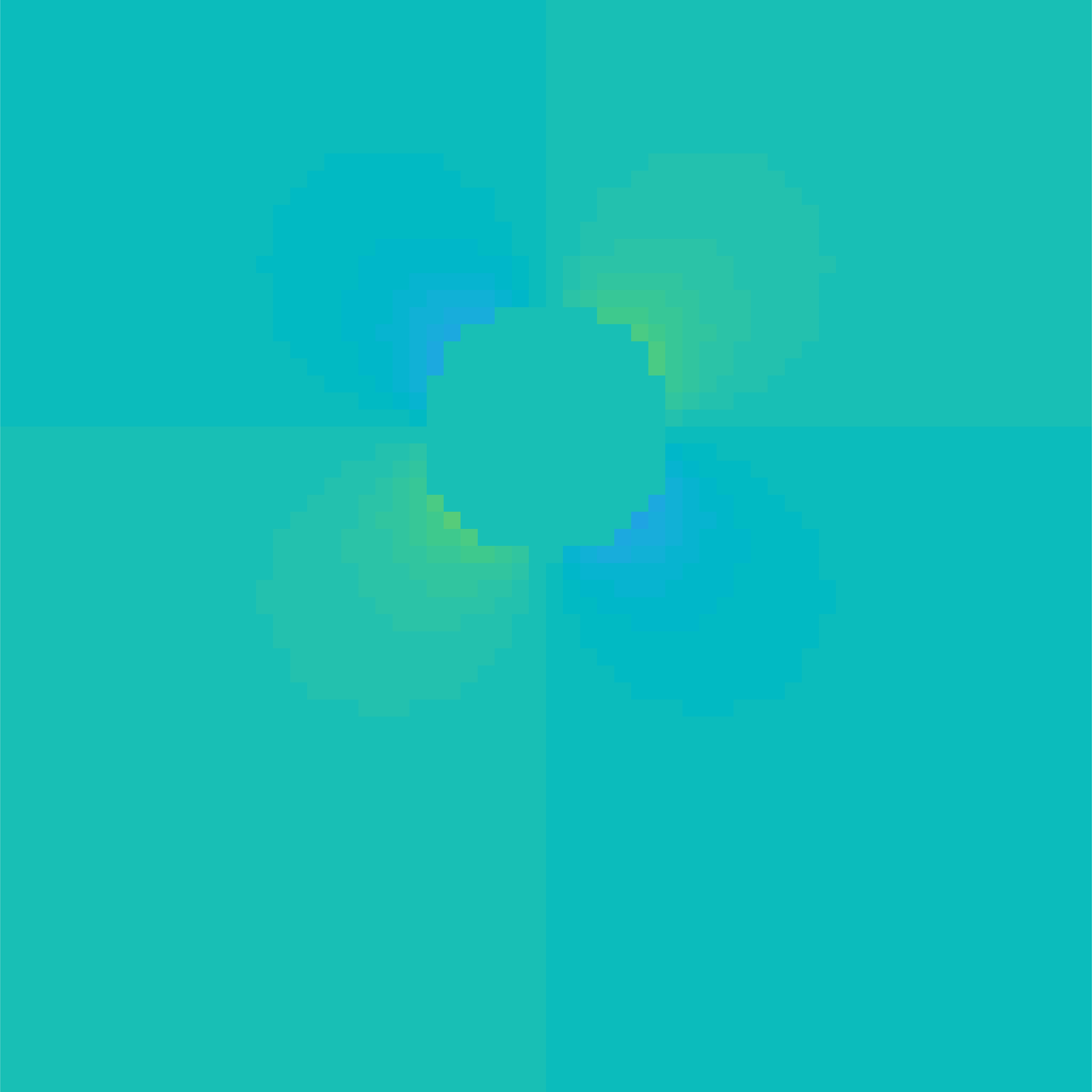}}~
\subfloat[Sequential \newline MSE=0.0030]{\includegraphics[width=0.2\textwidth]{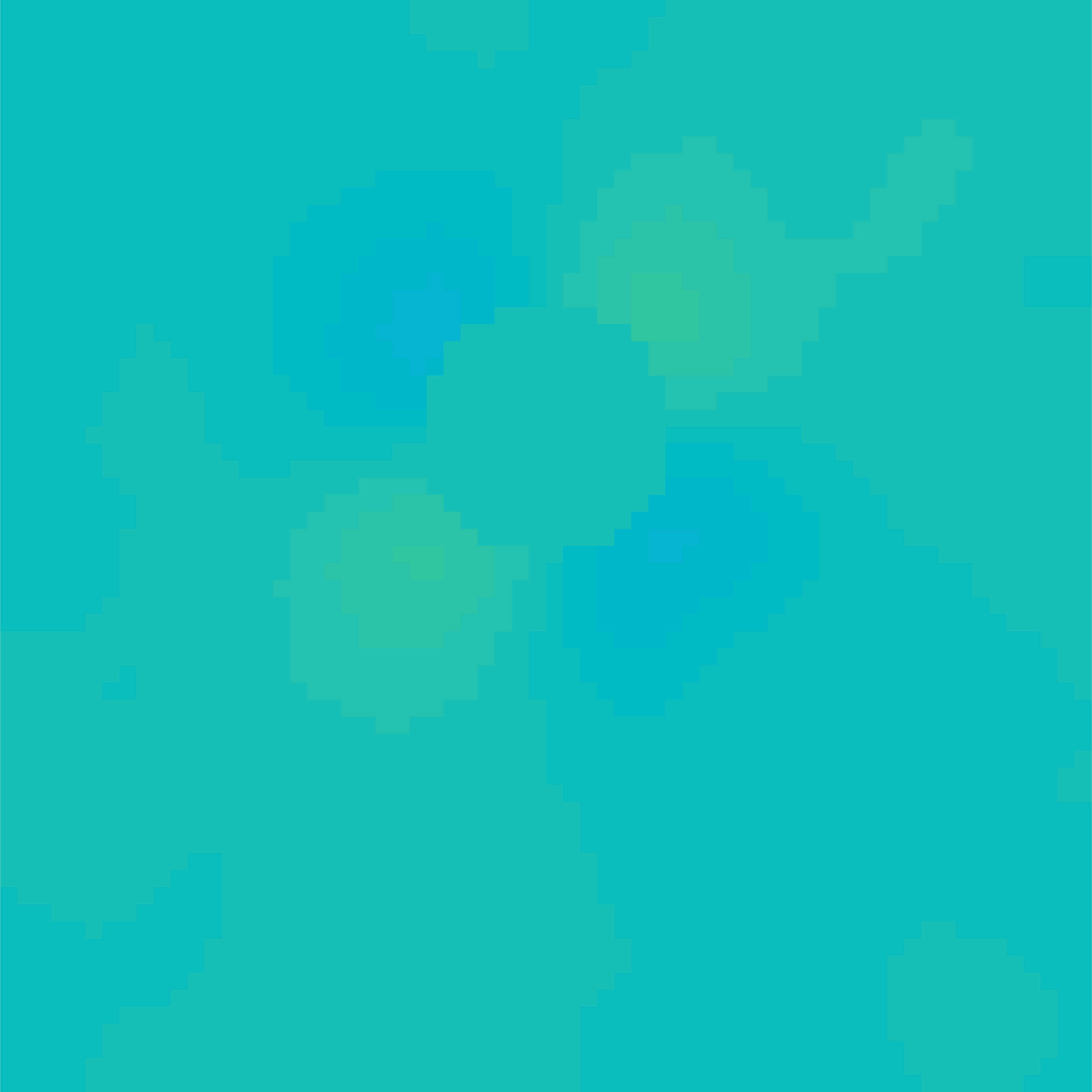}}~
\subfloat[Joint \newline MSE=0.0020]{\includegraphics[width=0.2\textwidth]{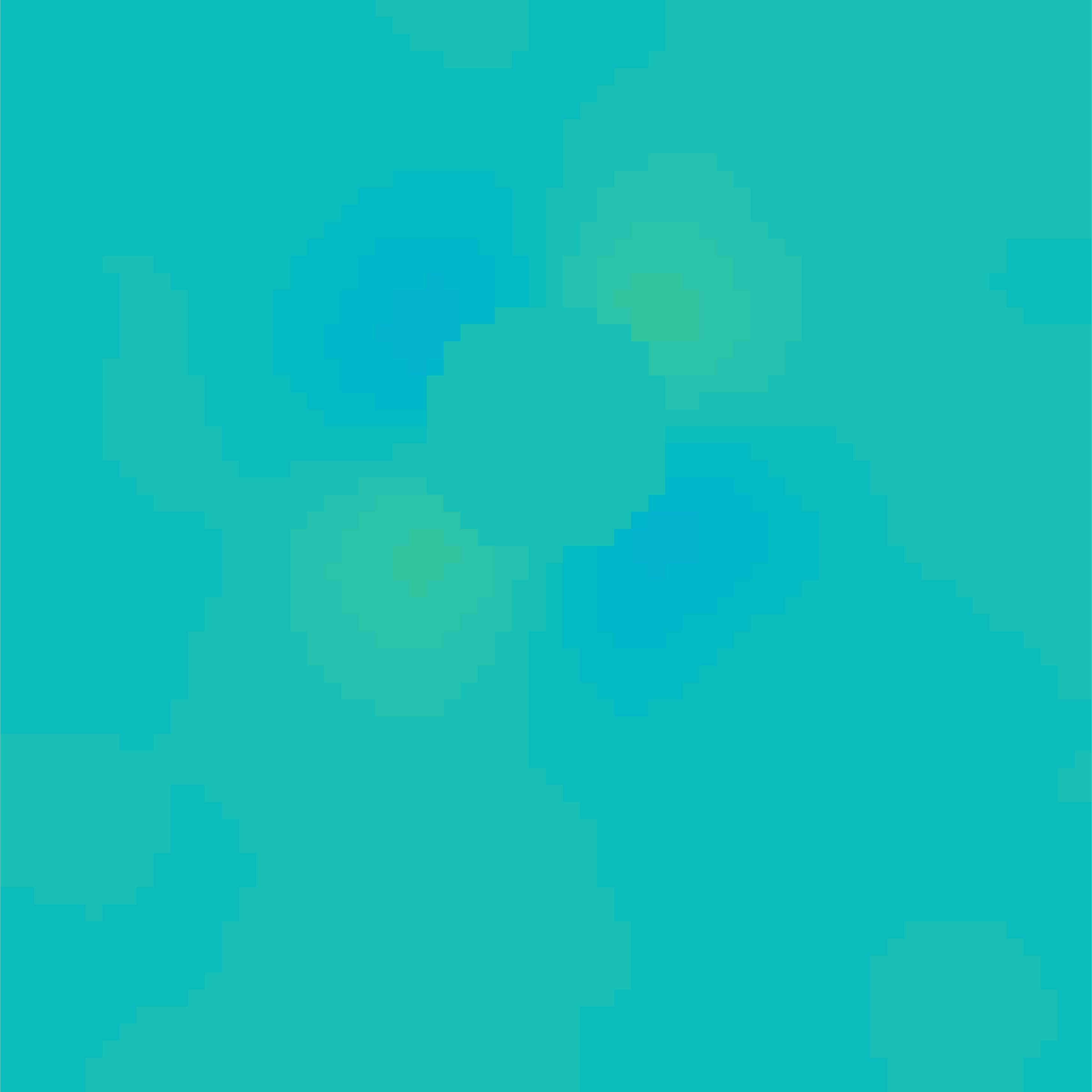}}

\subfloat[Groundtruth]{\includegraphics[width=0.2\textwidth]{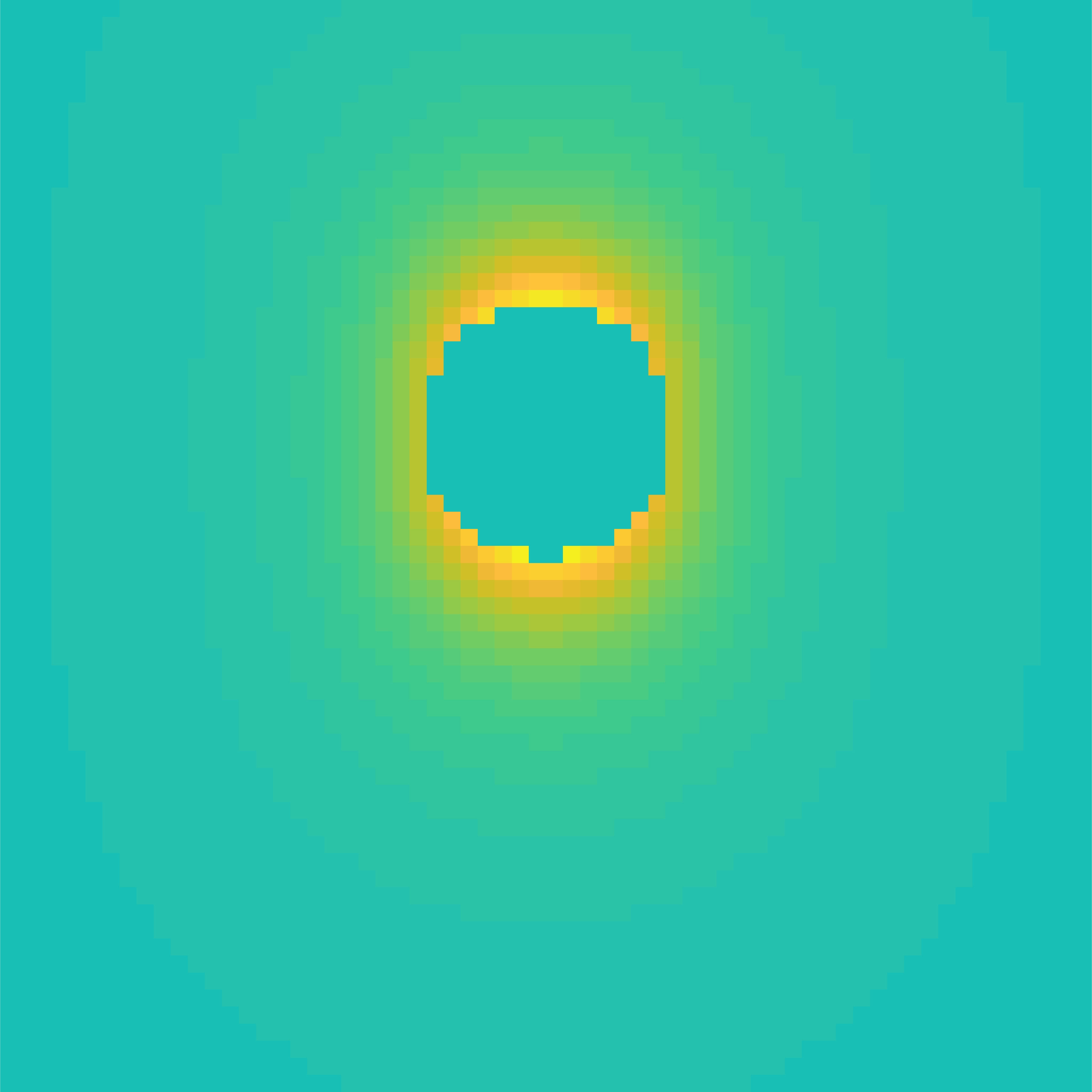}}~
\subfloat[Sequential \newline MSE=0.0046]{\includegraphics[width=0.2\textwidth]{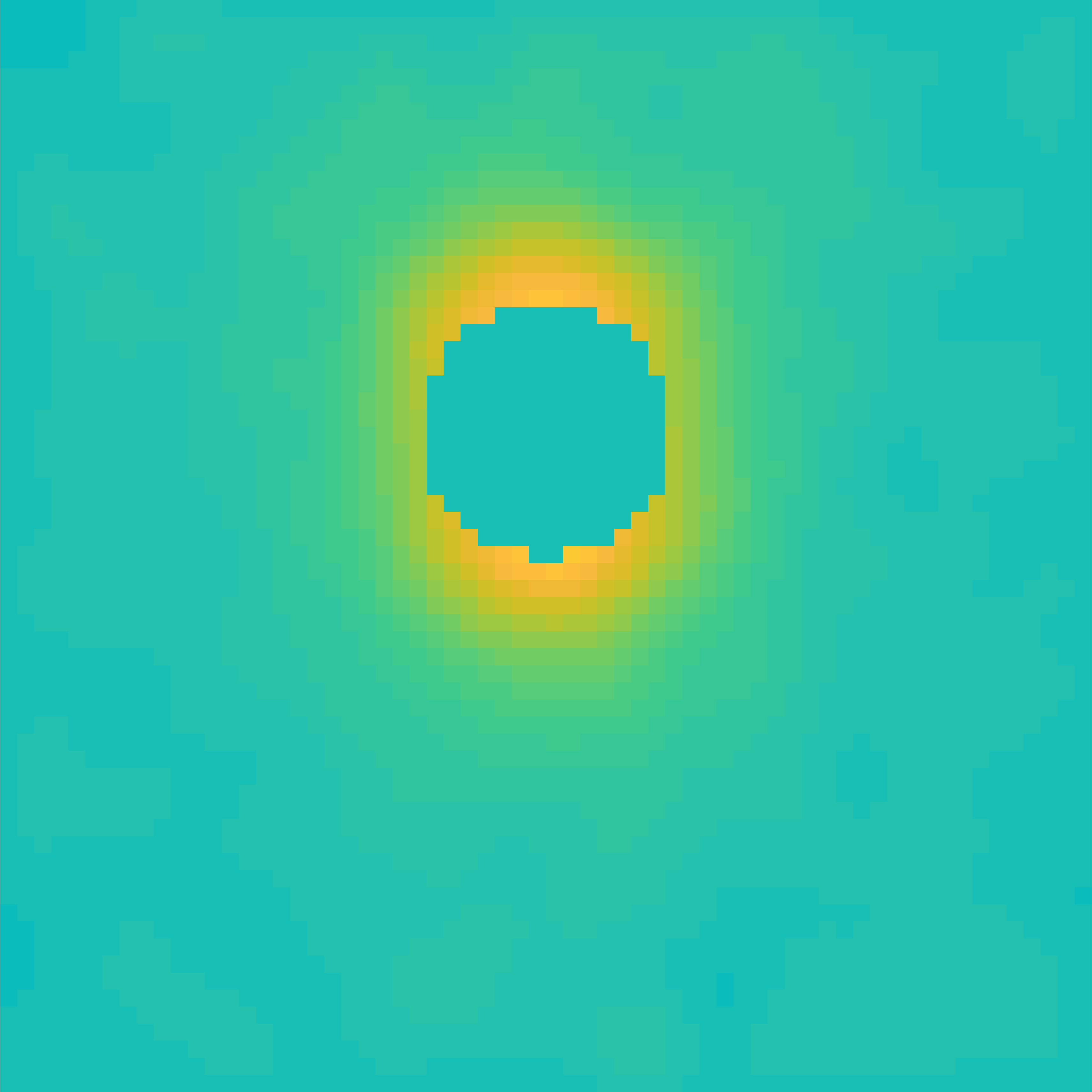}}~ 
\subfloat[Joint \newline MSE=0.0035]{\includegraphics[width=0.2\textwidth]{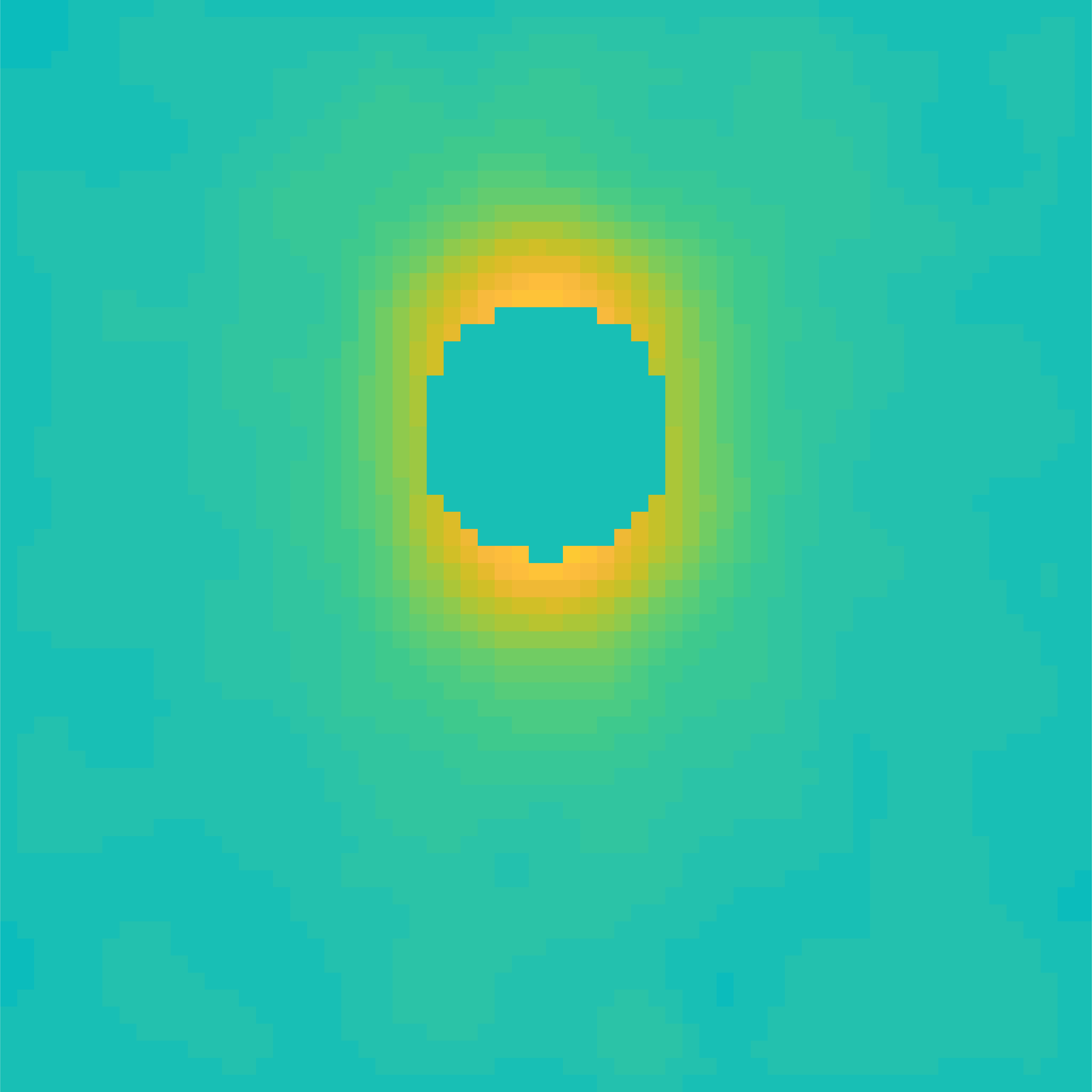}}

\subfloat[Groundtruth]{\includegraphics[width=0.2\textwidth]{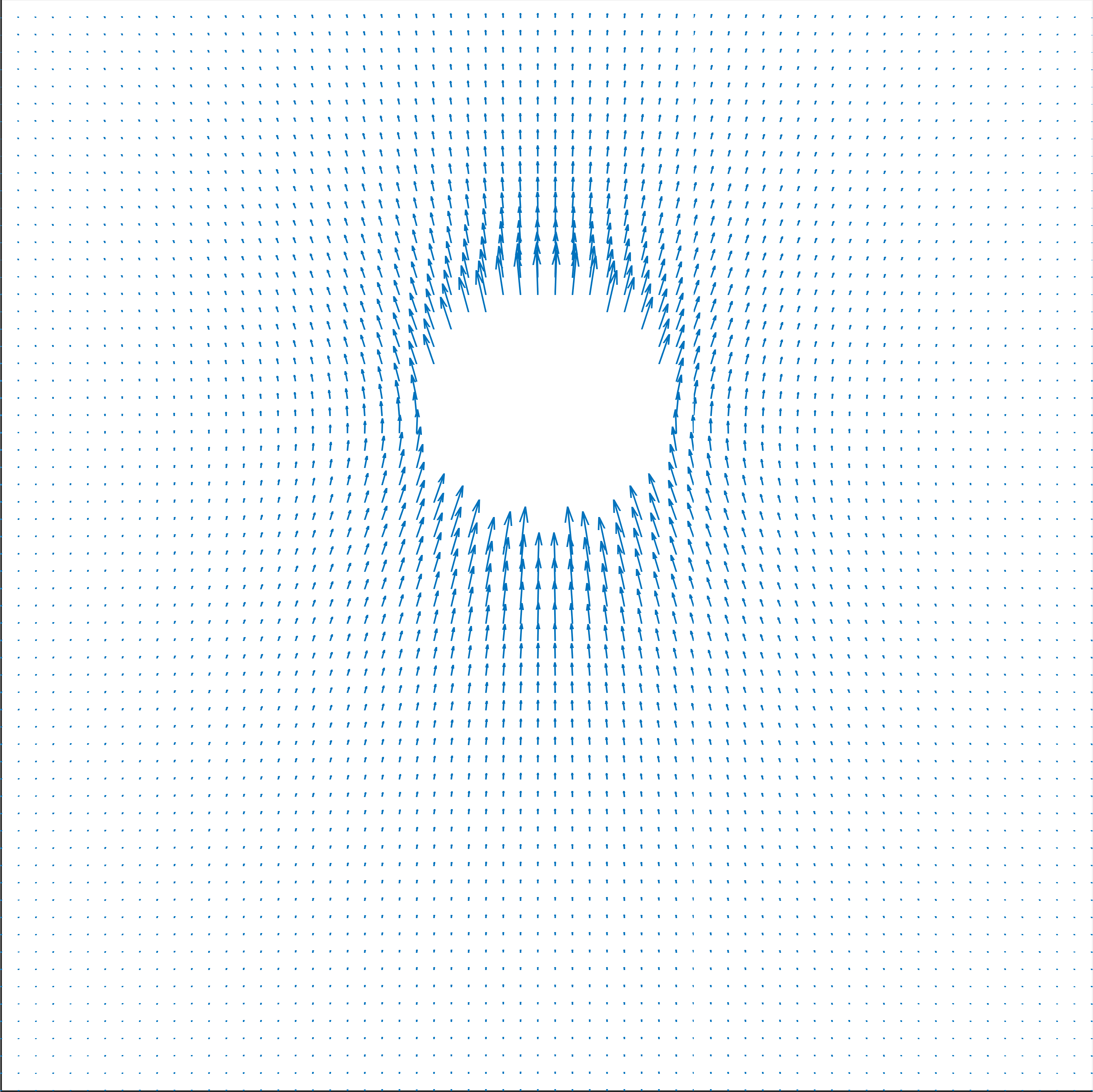}}~
\subfloat[Sequential]{ \includegraphics[width=0.2\textwidth]{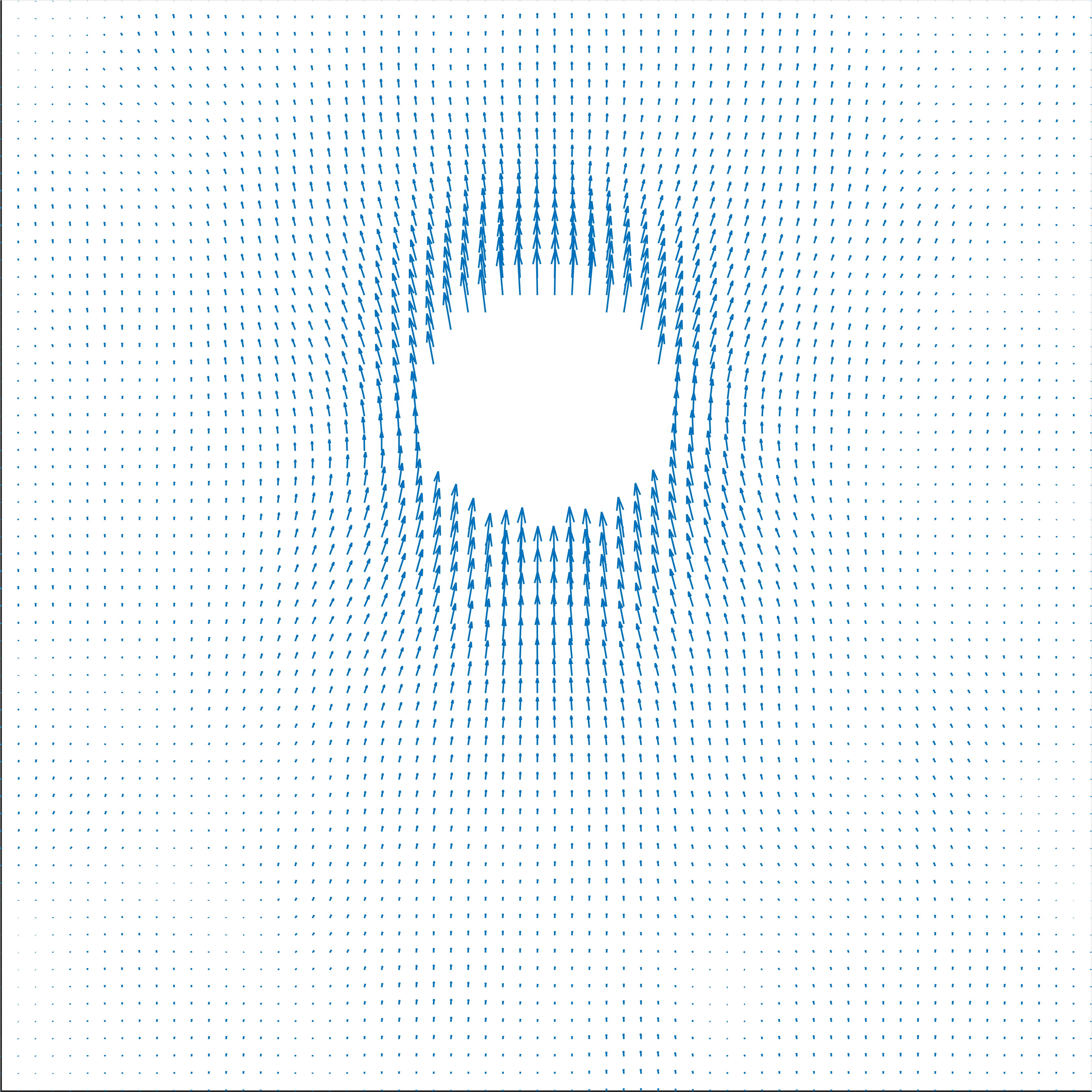}}~
\subfloat[Joint]{\includegraphics[width=0.2\textwidth]{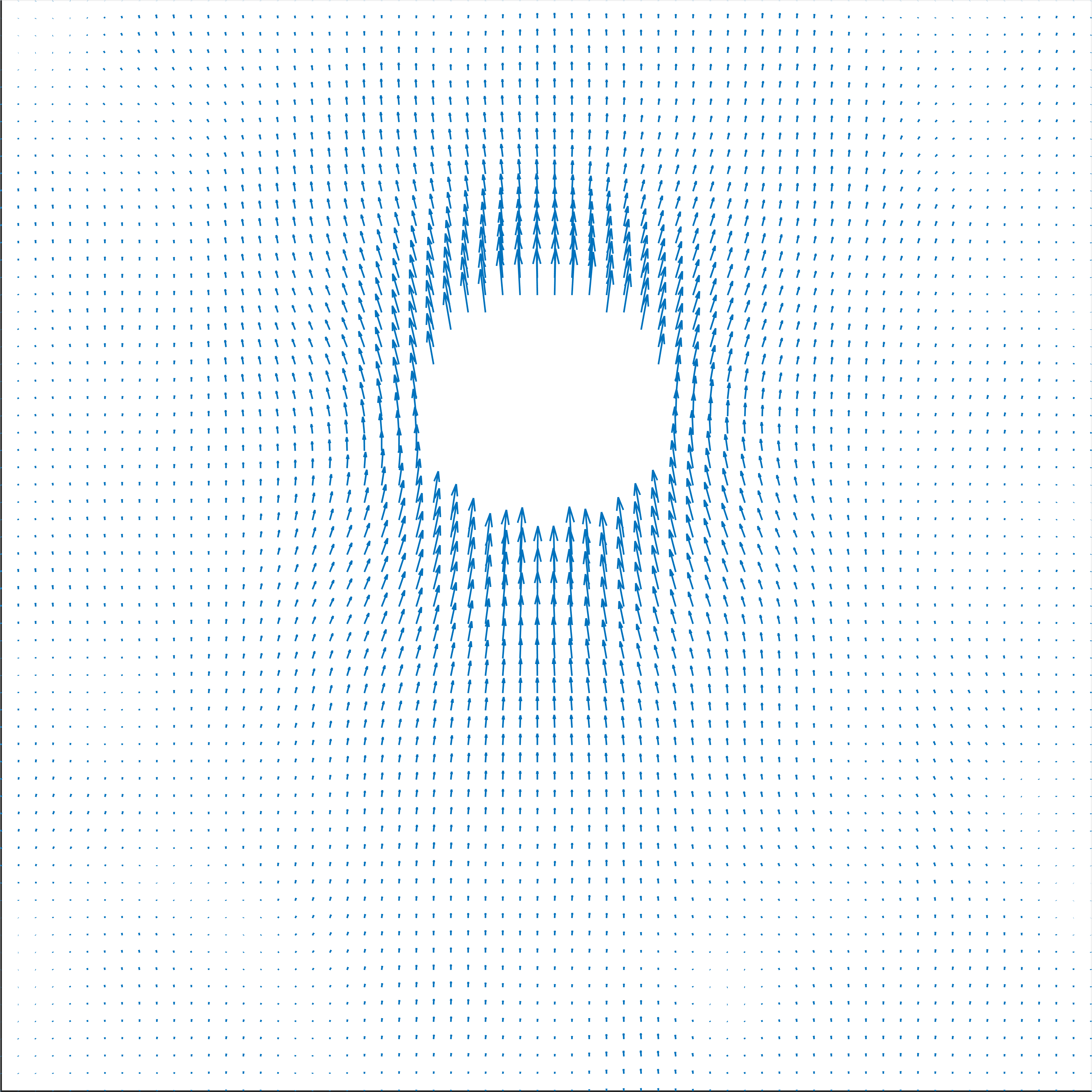}}

\caption{Phase reconstructions for the sequential approach and our joint approach compared to the ground truth. Top row: $x$ direction, middle row: $z$ direction, bottom row: velocity plots. We sampled 11\% of the k-space data.}
\label{fig:frame19}
    \end{figure}

In Fig.~\ref{fig:frame19} we can see the results for the sequential approach compared to the joint approach when sampling only 11\% of the k-space data. Although visually there is not significant change, the MSE shows a big improvement for the joint approach. This confirms that using our joint model is relevant for the problem of velocity-encoded MRI. For the 32 frames, we report the average MSE for magnitude and phase in Table~\ref{tab:1} where can see a drastic improvement compared to the sequential approach. 
\begin{table}[h!]
\caption{MSE for phase ($\phi_1$ and $\phi_2$) and magnitude ($u_1$ and $u_2$) images for the sequential and joint approaches. The error is significantly decreased using our proposed joint approach.}
\label{tab:1}   
\centering
\begin{tabular}{p{2cm}p{2cm}p{2cm}p{2cm}p{2cm}}
\hline\noalign{\smallskip}
 & $u_1$ & $u_2$ &  $\phi_1$ & $\phi_2$ \\ %$v_1$ & $v_2$ &
\noalign{\smallskip}\svhline\noalign{\smallskip}
%\noalign{\smallskip}\svhline\noalign{\smallskip}
Sequential & 0.0019 & 0.0028 & 0.0032 & 0.0059 \\ 
Joint & \textbf{0.0011}& \textbf{0.0012} & \textbf{0.0018} & \textbf{0.0051} \\
\noalign{\smallskip}\hline\noalign{\smallskip}
\end{tabular}
\end{table}

%\begin{table}
%\begin{tabular}{p{2cm}p{2.4cm}p{2cm}p{4.9cm}}
%\hline\noalign{\smallskip}
%Classes & Subclass & Length & Action Mechanism  \\
%\noalign{\smallskip}\svhline\noalign{\smallskip}
%Translation & mRNA$^a$  & 22 (19--25) & Translation repression, mRNA cleavage\\
%Translation & mRNA cleavage & 21 & mRNA cleavage\\
%Translation & mRNA  & 21--22 & mRNA cleavage\\
%Translation & mRNA  & 24--26 & Histone and DNA Modification\\
%\end{tabular}
%$^a$ Table foot note (with superscript)
%\vspace*{-12pt}
%\end{table}
\subsection{Real dataset}
In this section we present our model performance on real data acquired with the following protocol described in \cite{Reci} and briefly reported here. 
\paragraph{Acquisition protocol} The experiments were conducted on an AV-400 Bruker magnet, operating at a resonant frequency of 400.25 MHz for ${}^{1}$H observation with an RF coil of 25 mm diameter. The maximum magnetic field gradient amplitude available in each spatial direction is 146 G$\text{cm}^{-1}$. 
The velocity images were acquired with a 2D MR spiral imaging technique developed and published elsewhere \cite{TaylerHollandSedermanEtAl2012}. %, shown in Fig.~\ref{fig:exp}. 
%Images are of 64 $\times$ 64 pixels and of resolution $265 \mu$ m $\times 265 \mu $m, %The 2D images are acquired over a slice thickness of 150 $\mu$ m. 1024 complex points spaced at 2 $\mu$s are 
%acquired along the spiral, corresponding to 25\% sampling percentage and imaging time of 2.05 ms. 
Images were acquired with 64 $\times$ 64 pixels over a field of view of 17 mm $\times$ 17 mm resulting in an image resolution of 265 mm $\times$ 265 mm. Data in k-space were acquired along a spiral trajectory at a sampling rate corresponding to 25\% of full Nyquist sampling over a time of 2.05 ms for the entire image.
%The spiral density and coverage of k-space is optimised according the method outlined in Section 9.3.1; the optimal spiral density is close to being constant and the spiral stretches to ~ 40\% of kmax, as illustrated in Fig, 9.1(b). The velocity-encoding gradients are designed with $\delta = 0.26 $ ms, $\Delta$ = 0.49 ms and g = 29 G cm$^{-1}$. The RF excitation pulse duration is 200 $\mu s$ and is designed for a flip angle of 11$^{\circ}$, to allow for a fast return to equilibrium of the magnetization vector. The timings of the RF pulse, velocity-encoding gradients and imaging gradients allow for a repeat of the pulse sequence every 4.0 ms.

\begin{figure}[t]
\centering
\includegraphics[width=\textwidth]{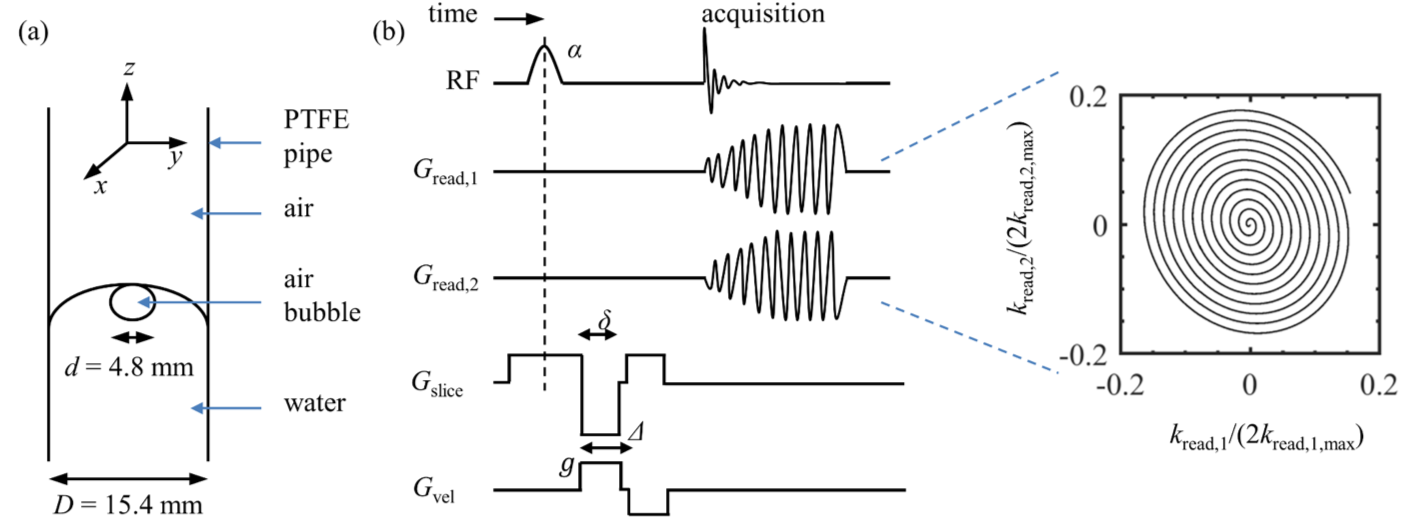}
\caption{(a) Schematic of experimental setup. (b) Pulse sequence used for MR velocimetry acquisitions and the corresponding k-space traversal. Taken from \cite{Reci}.}
\label{fig:exp}
\end{figure}

We acquire the three velocity components for a transverse slice (perpendicular to the axis of the pipe) and a longitudinal slice (parallel to the axis of the pipe), cutting through approximately the centre of the bubble.
%Three component velocity maps on a 2D image are obtained for a transverse slice (perpendicular to the axis of the pipe) and a longitudinal slice (parallel to the axis of the pipe). Both slices cut through approximately the centre of the bubble. 
For a given slice direction (transverse or longitudinal) and a given direction of the velocity, four measurements corresponding to the application of the velocity-encoding gradient with alternating polarity and to the flow compensation, are taken, as discussed in Sect.~\ref{sec:velocityMRI} (see Fig.~\ref{fig:exp}(b)). %to be measured ($x$, $y$ or $z$), the one component velocity map ($v_x$, $v_y$ or $v_z$) is acquired by applying repeatedly every 4 ms %the pulse sequence in Fig.~\ref{fig:exp}(b) 
%with the velocity-encoding gradient in the respective direction ($x$, $y$ or $z$) and with alternating polarity between consecutive pulse sequences (from $\pm g$ to $\mp g$). 
The final velocity for each component is then obtained as the difference between the phase of the MRI images reconstructed from the acquired k-space data of consecutive pulse sequences with flow on, and the reference to the zero flow experiment (see Sections \ref{sec:flowphase} and \ref{sec:zeroflow}, respectively). %, the one component velocity map at 4 ms temporal resolution is reconstructed, as discussed in Sect.~\ref{sec:2}. %The maps of the other components of velocity are obtained in a similar way at temporal resolutions of 4 ms but with the pulse sequences applied to different bursting bubbles. The size, position and the hydrodynamics of different bursting bubbles at the free surface is consistent. As a result, the one component velocity maps in different directions, $v_x$, $v_y$ or $v_z$, are superimposed, by aligning the point in time at which the bubble burst starts, to effectively give three component velocity maps at a temporal resolution of 4 ms. 
% This is from Andi's thesis!!

\begin{figure}[t]
   \centering
\subfloat[Zero-filled]{\includegraphics[width=0.3\textwidth]{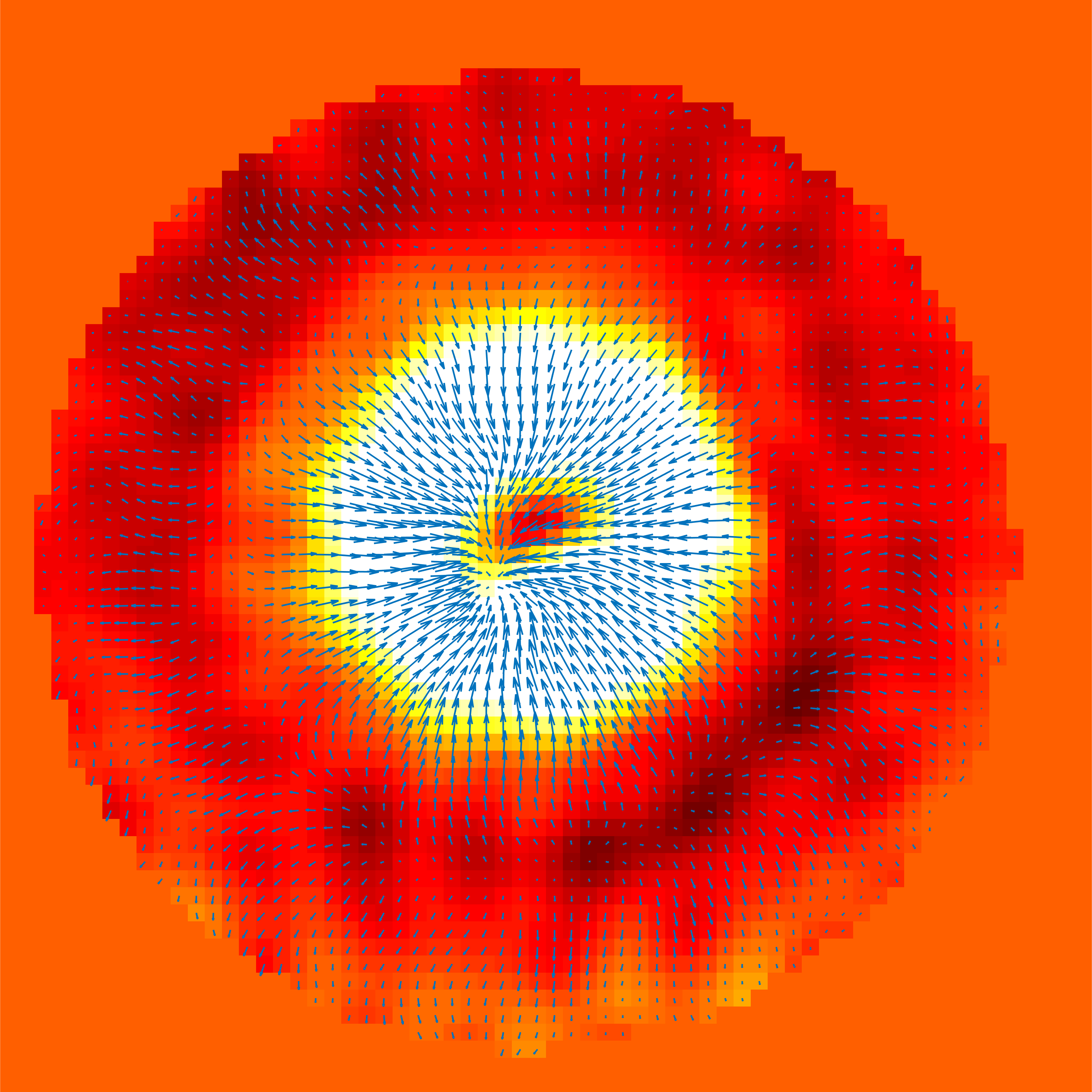}}~
\subfloat[Sequential]{\includegraphics[width=0.3\textwidth]{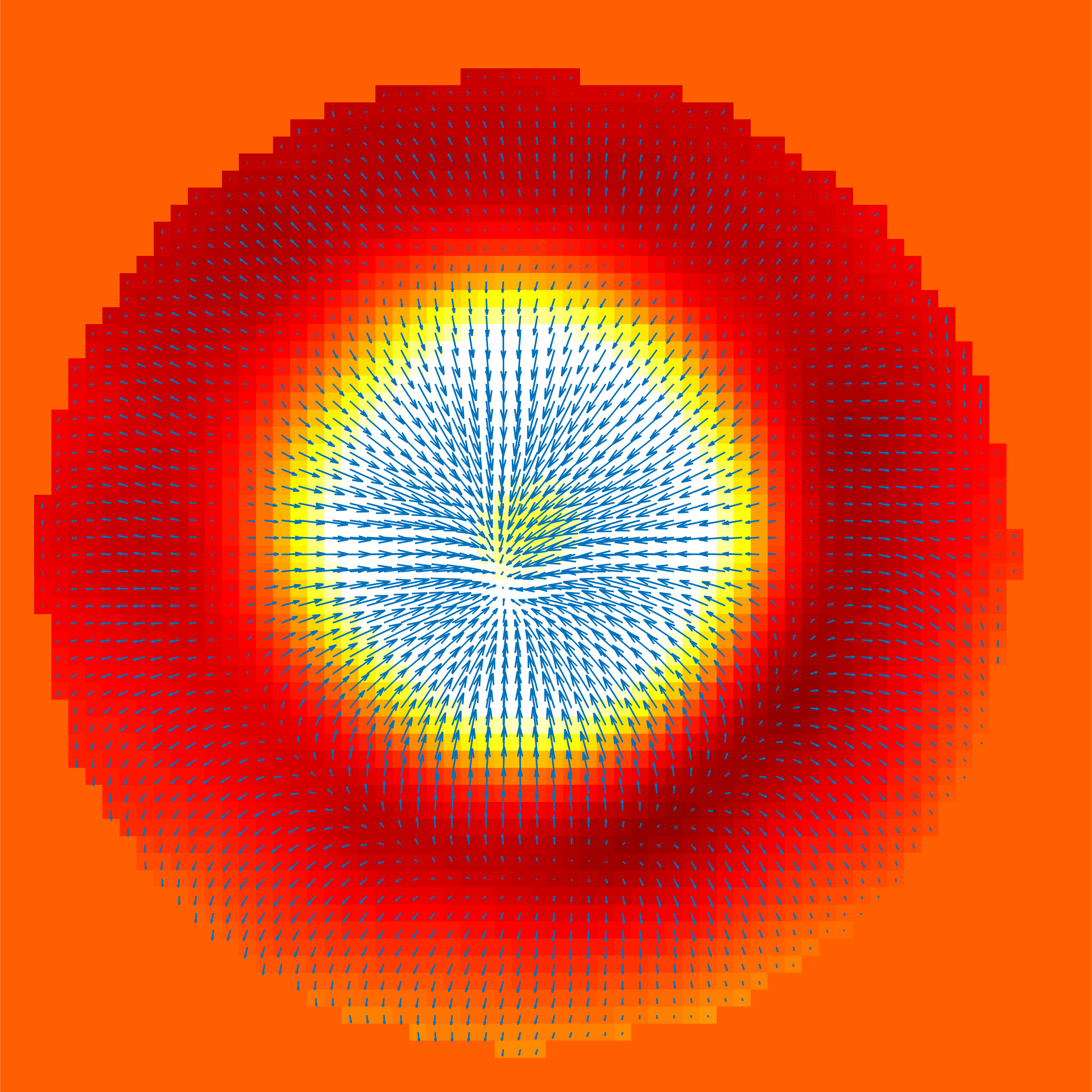}}~
\subfloat[Joint ]{\includegraphics[width=0.34\textwidth,]{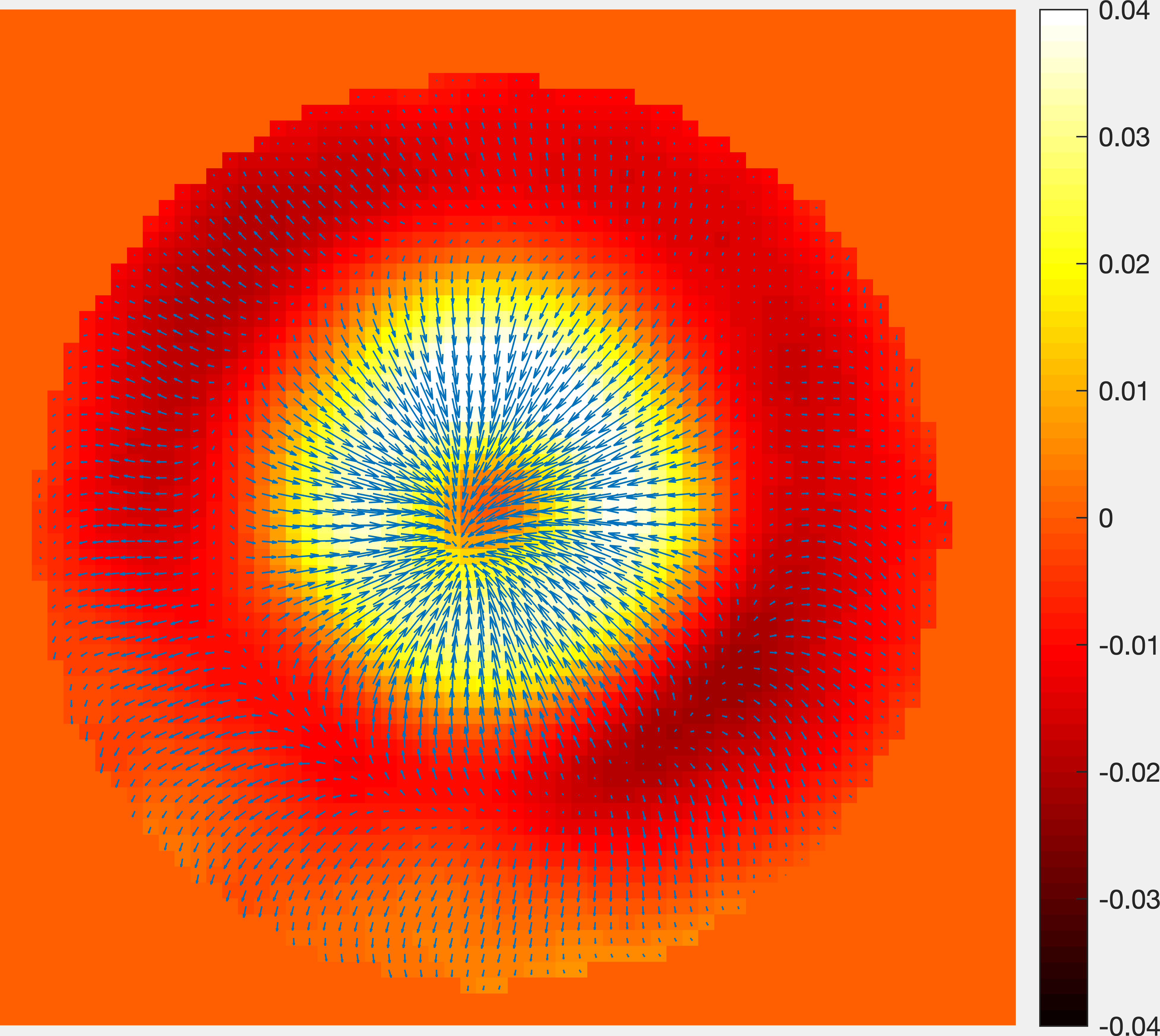}}
%\caption{Phase reconstructions for the sequential approach and our joint approach compared to the zero-filling solution. }
%\label{fig:frame4trans}

   \centering
\subfloat[Zero-filled]{\includegraphics[width=0.3\textwidth]{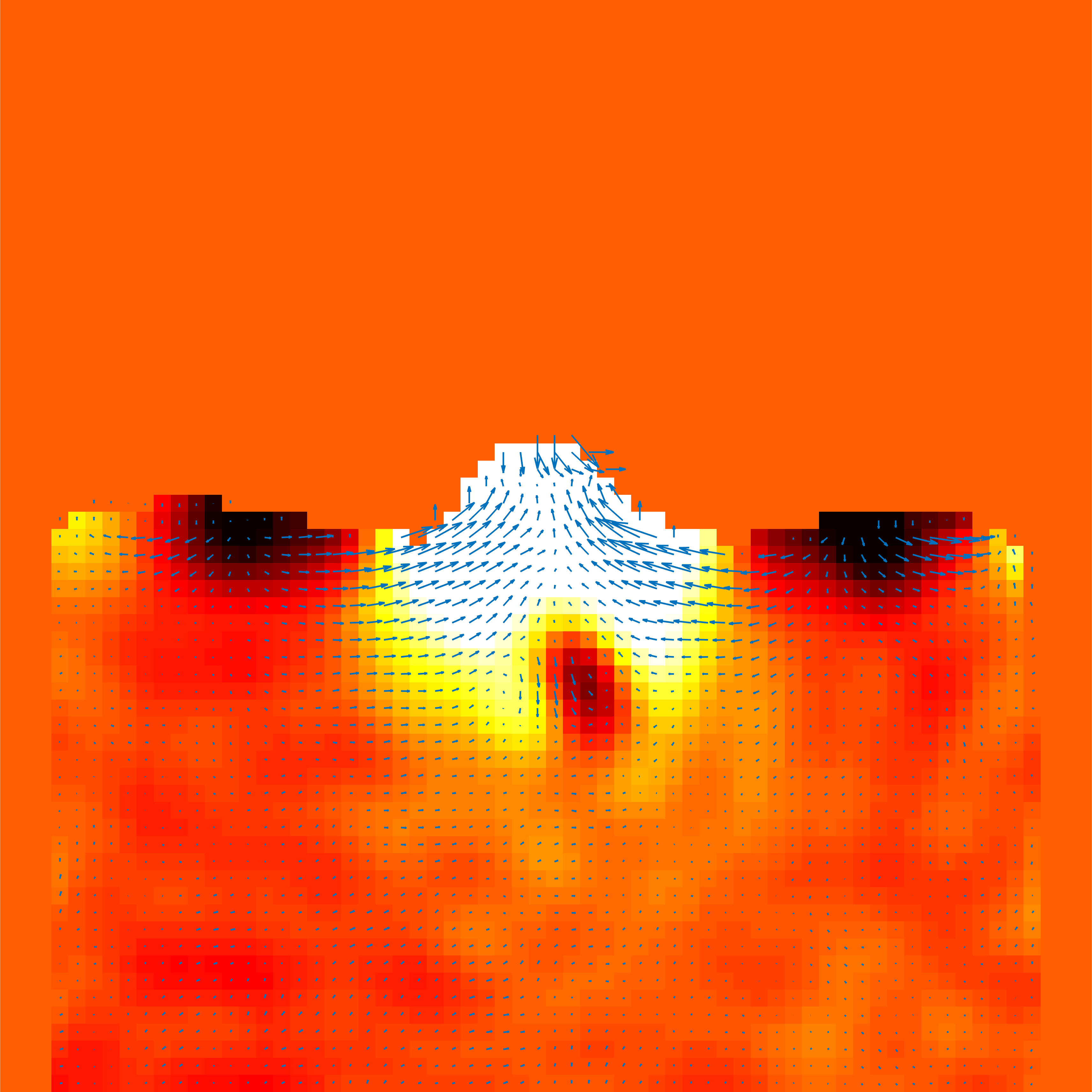}}~
\subfloat[Sequential  \label{fig:long_seq}]{\includegraphics[width=0.3\textwidth]{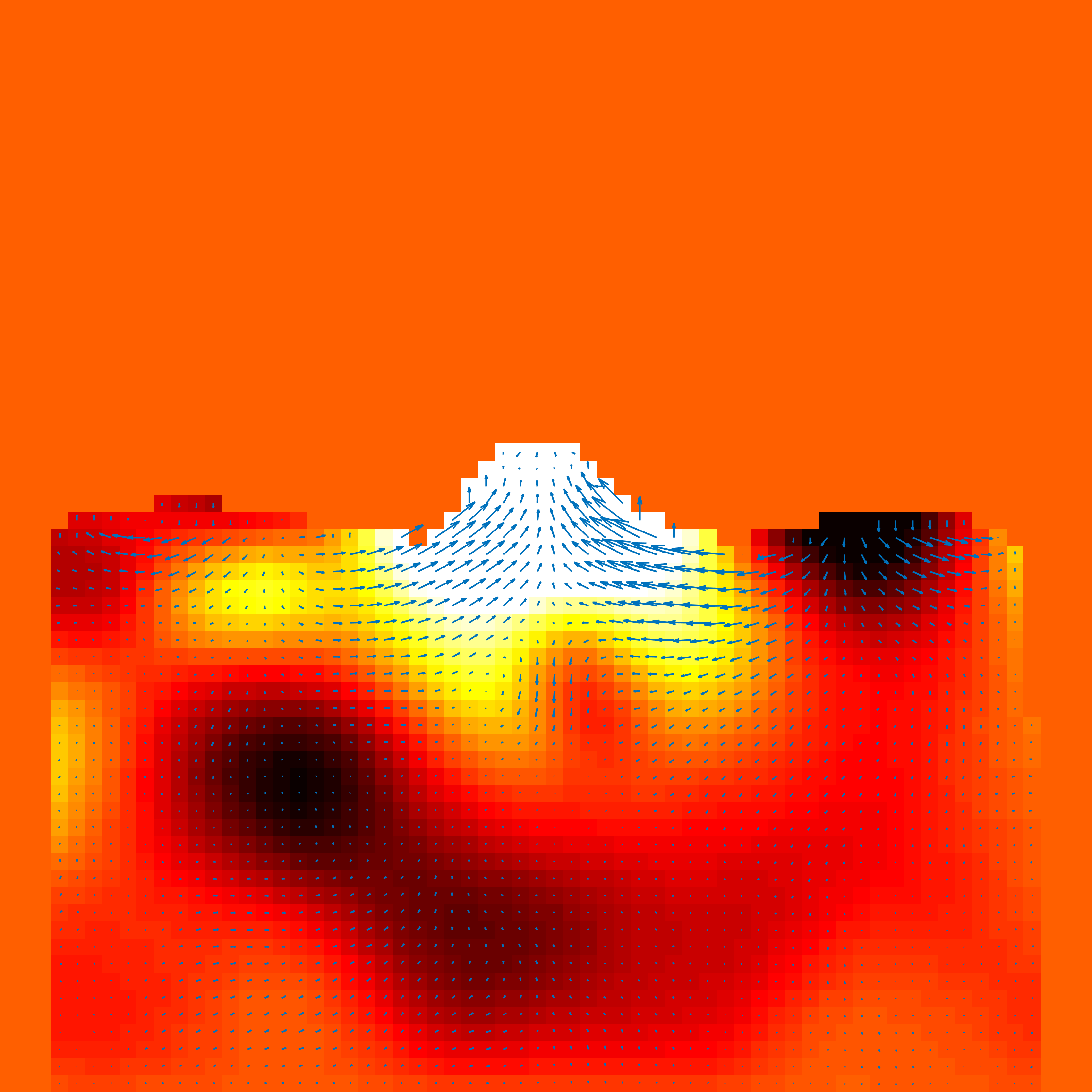}}~
\subfloat[Joint ]{\includegraphics[width=0.34\textwidth,]{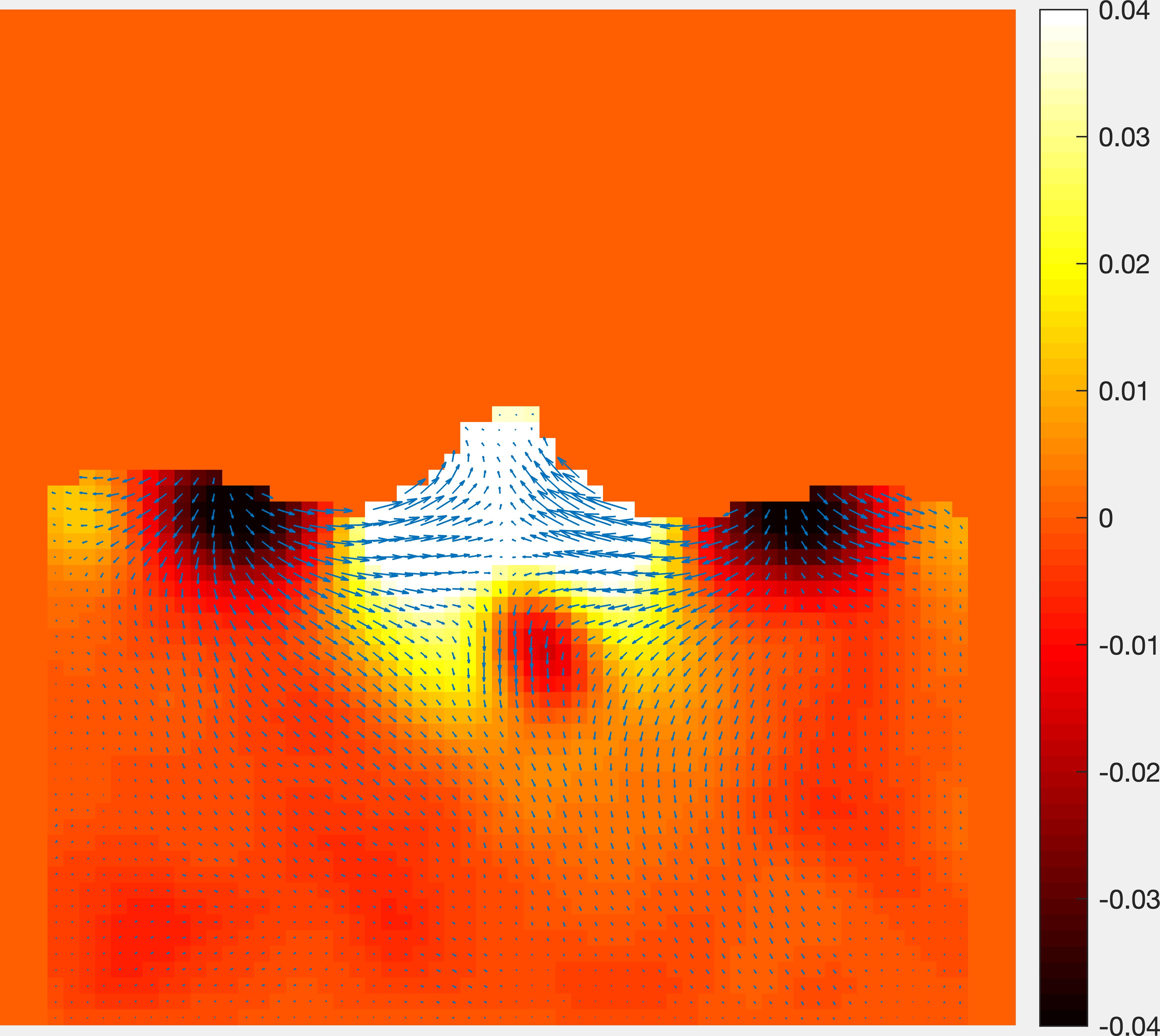}}
\caption{Phase reconstructions for the sequential approach and our joint approach compared to the zero-filling solution. Results for a bursting bubble from a transversal view (top row) and longitudinal view (bottom row).}
\label{fig:frame4}
    \end{figure}

\begin{figure}[h]
   \centering
\subfloat[ Zero-filled]{\includegraphics[width=0.3\textwidth]{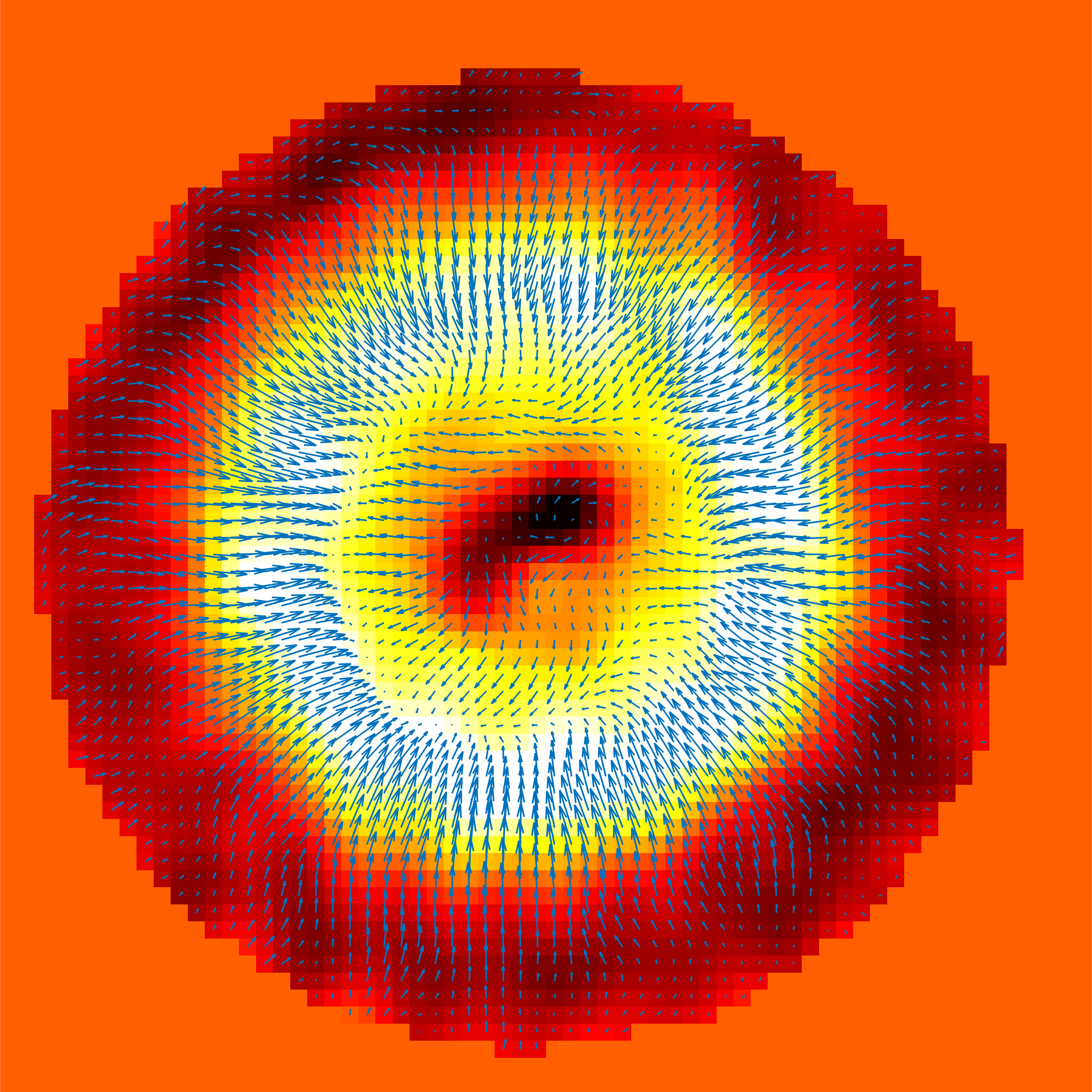}}~
\subfloat[Sequential ]{\includegraphics[width=0.3\textwidth]{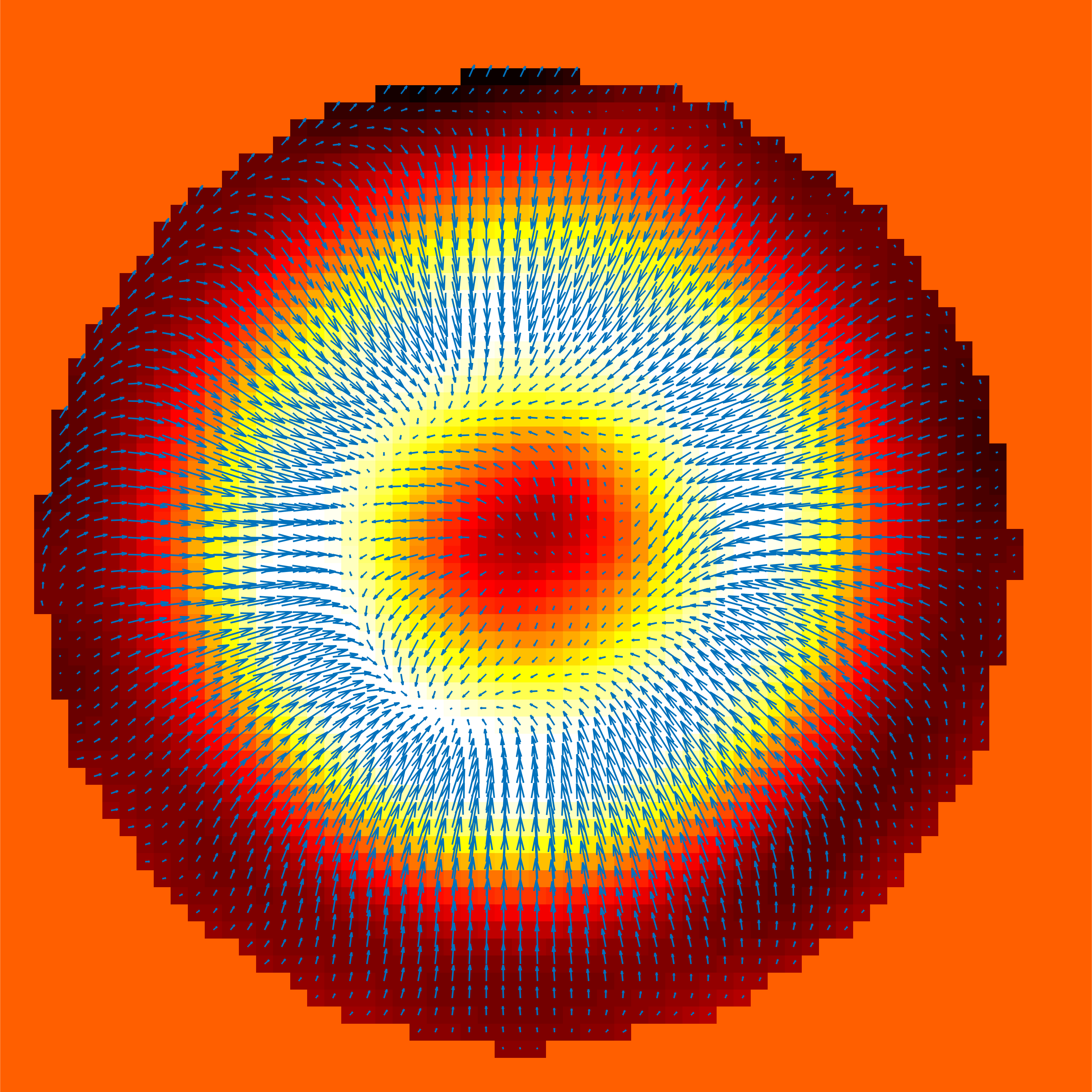}}~
\subfloat[Joint ]{\includegraphics[width=0.34\textwidth,]{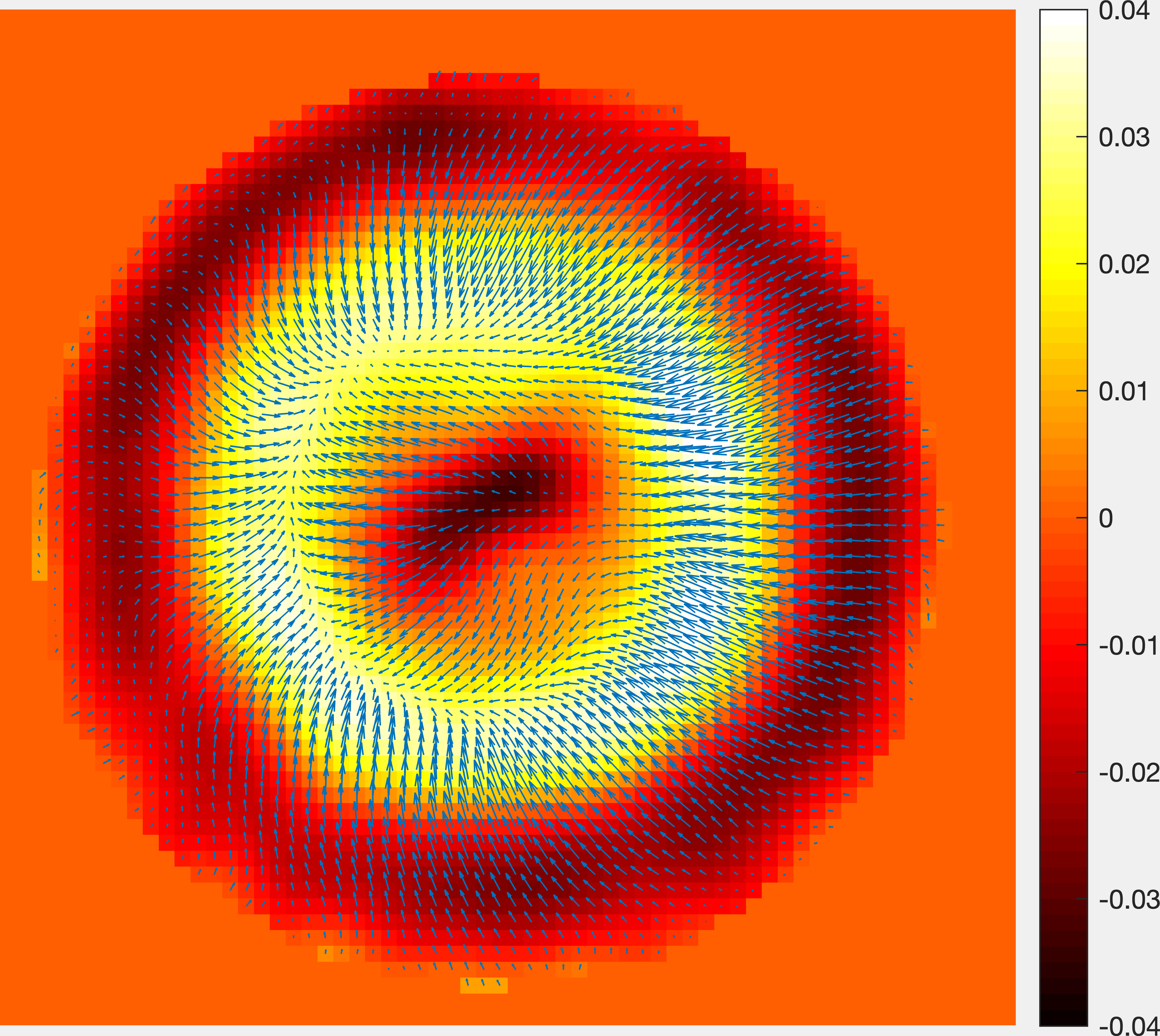}}
%\caption{Phase reconstructions for the sequential approach and our joint approach compared to the zero-filling solution. }
%\label{fig:frame6trans}

   \centering
\subfloat[ Zero-filled]{\includegraphics[width=0.3\textwidth]{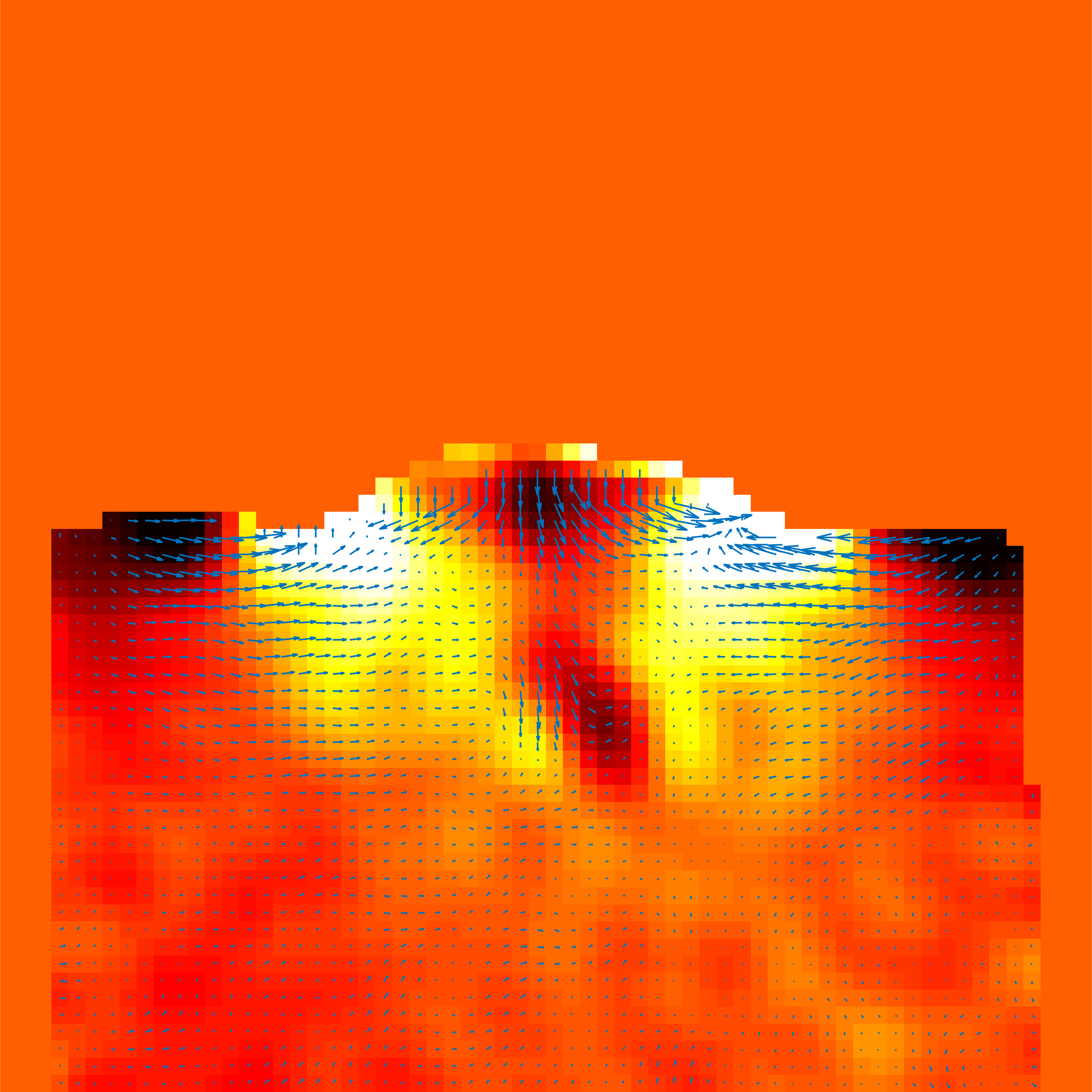}}~
\subfloat[Sequential ]{\includegraphics[width=0.3\textwidth]{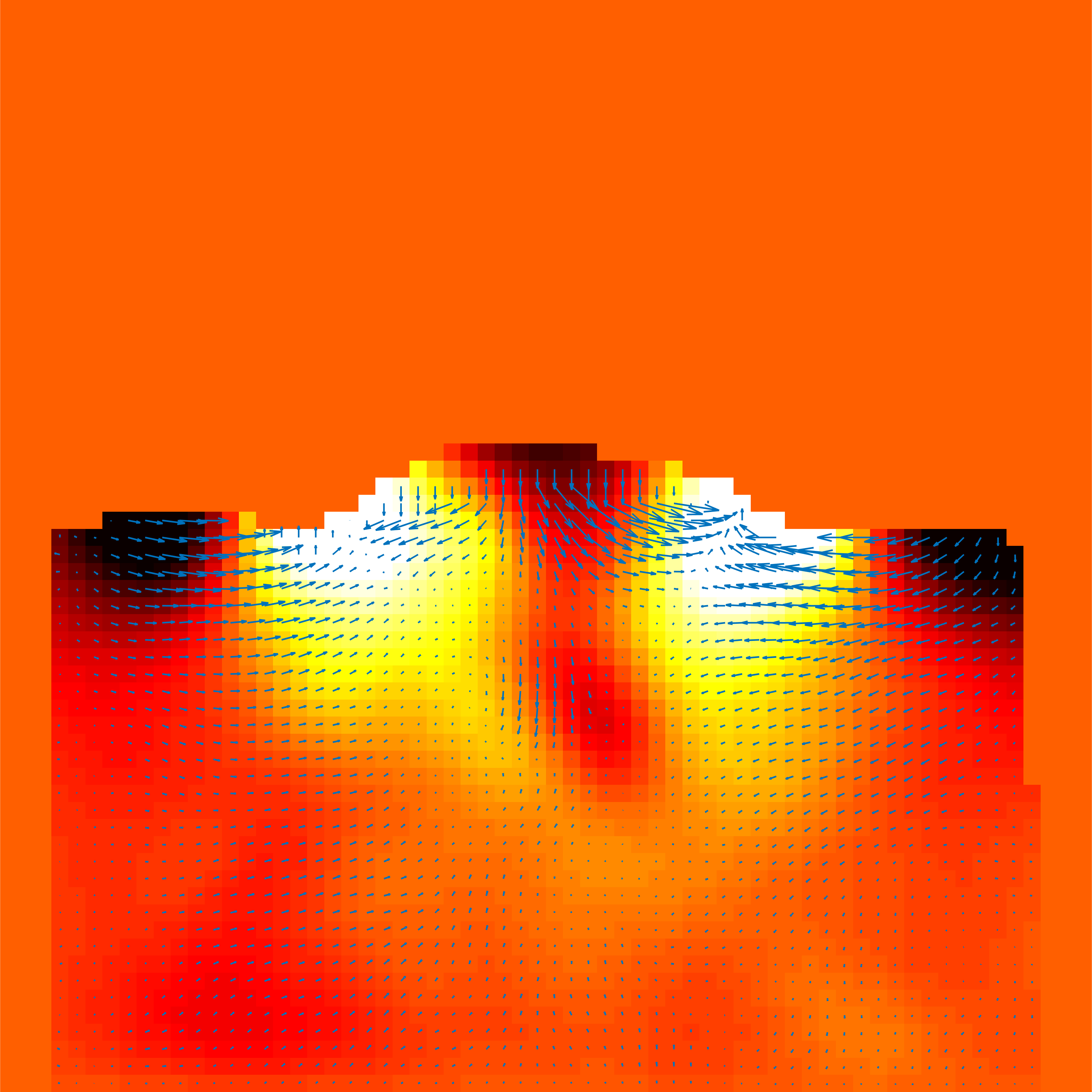}}~
\subfloat[Joint ]{\includegraphics[width=0.34\textwidth,]{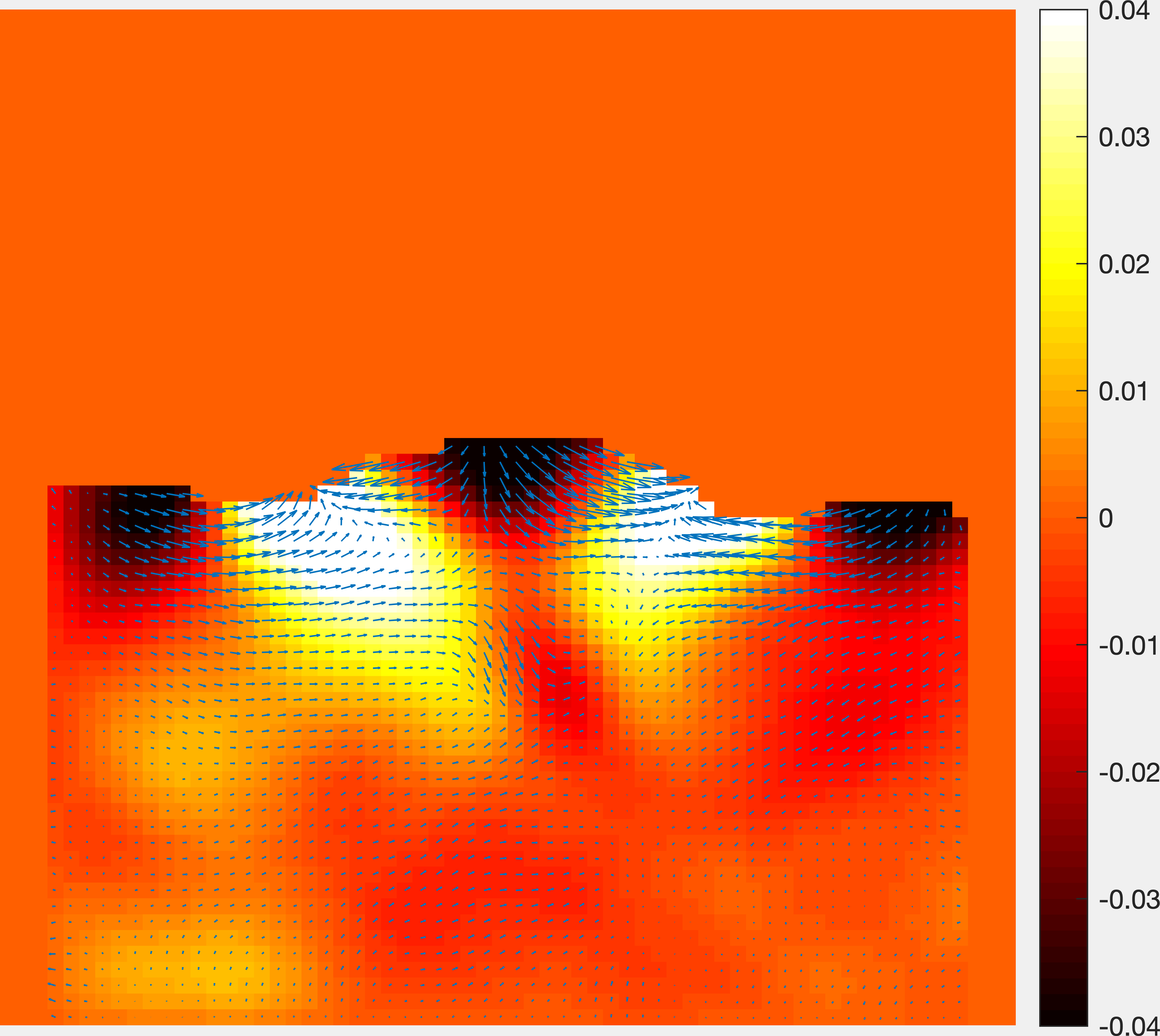}}
\caption{Phase reconstructions for the sequential approach and our joint approach compared to the zero-filling solution. Top row: transversal view. Bottow row: longitudinal view.}
\label{fig:frame6}
    \end{figure}

\paragraph{Experimental results on real data}
For the real data acquired with protocol described above, we present the results for our joint model in comparison with the zero-filling solution and the corresponding sequential approach also used in the previous subsection. In Fig.~\ref{fig:frame4} we show the result for a specific time frame for a bubble in a transversal and longitudinal view. At this specific time, the bubble is bursting which corresponds to an upward jet being ejected.  As we can see, the zero-filling solution gives an indication of the flow velocity but it is very noise and imprecise. In contrast, the joint approach removes noise and successfully estimate the velocity flow. The sequential approach on the other hand, although it produces a smoother reconstruction, results in small errors (see e.g. Fig.~\ref{fig:long_seq} on the left). In Fig.~\ref{fig:frame6} we observe similar results for a different time frame. We refer to the Appendix for the full dynamic sequence result.\\

\begin{figure}[h]
   \centering
\subfloat[Zero-filled]{\includegraphics[width=0.3\textwidth]{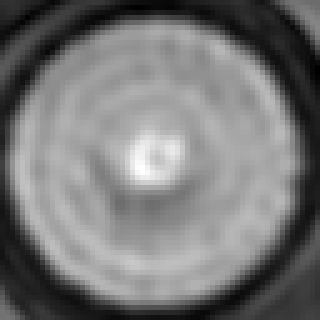}}~
\subfloat[Sequential ]{\includegraphics[width=0.3\textwidth]{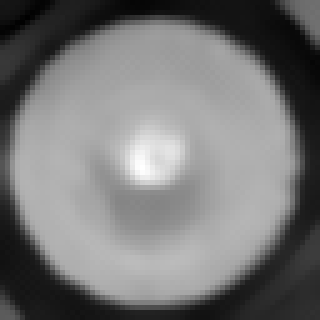}}~
\subfloat[Joint ]{\includegraphics[width=0.3\textwidth,]{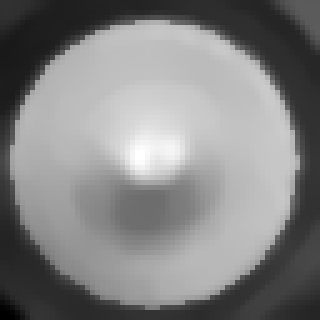}}

\subfloat[Zero-filled]{\includegraphics[width=0.3\textwidth]{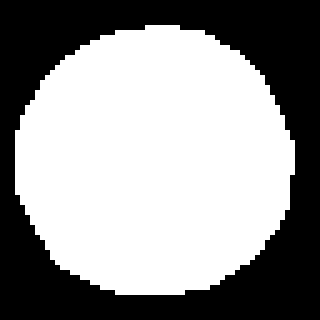}}~
\subfloat[Sequential ]{\includegraphics[width=0.3\textwidth]{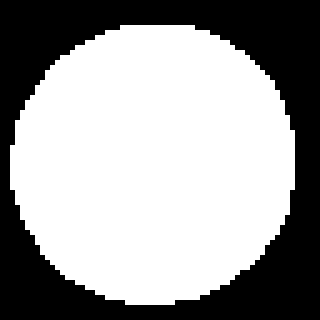}}~
\subfloat[Joint ]{\includegraphics[width=0.3\textwidth,]{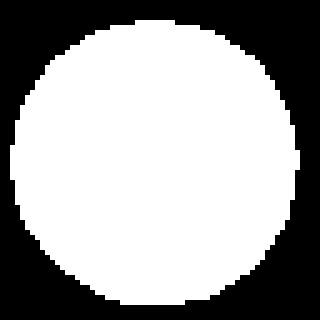}}

\caption{Magnitude reconstructions (top row) and corresponding segmentations (bottom row) for the sequential approach and our joint approach compared to the zero-filling solution. Transversal view.} %Results for a bursting bubble from a transversal view (top row) and longitudinal view (bottom row).}
\label{fig:frame4magn_trans}
    \end{figure}

\begin{figure}[h]
   \centering
\subfloat[Zero-filled]{\includegraphics[width=0.3\textwidth]{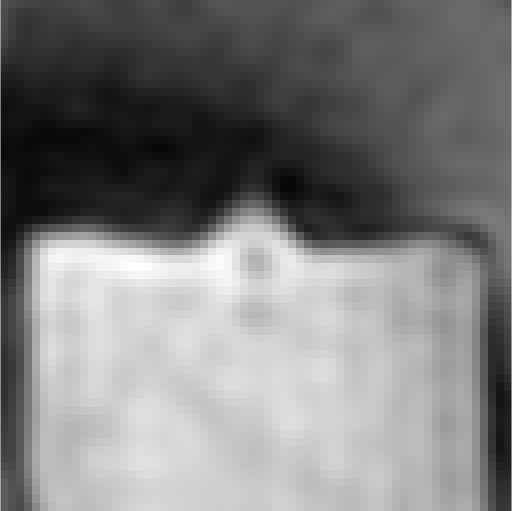}}~
\subfloat[Sequential ]{\includegraphics[width=0.3\textwidth]{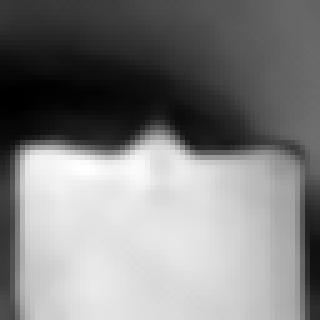}}~
\subfloat[Joint ]{\includegraphics[width=0.3\textwidth,]{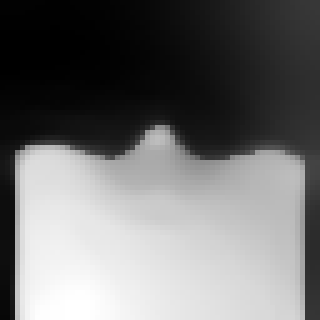}}

\subfloat[Zero-filled]{\includegraphics[width=0.3\textwidth]{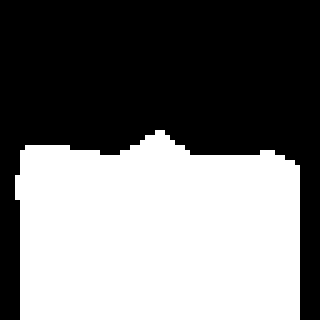}}~
\subfloat[Sequential ]{\includegraphics[width=0.3\textwidth]{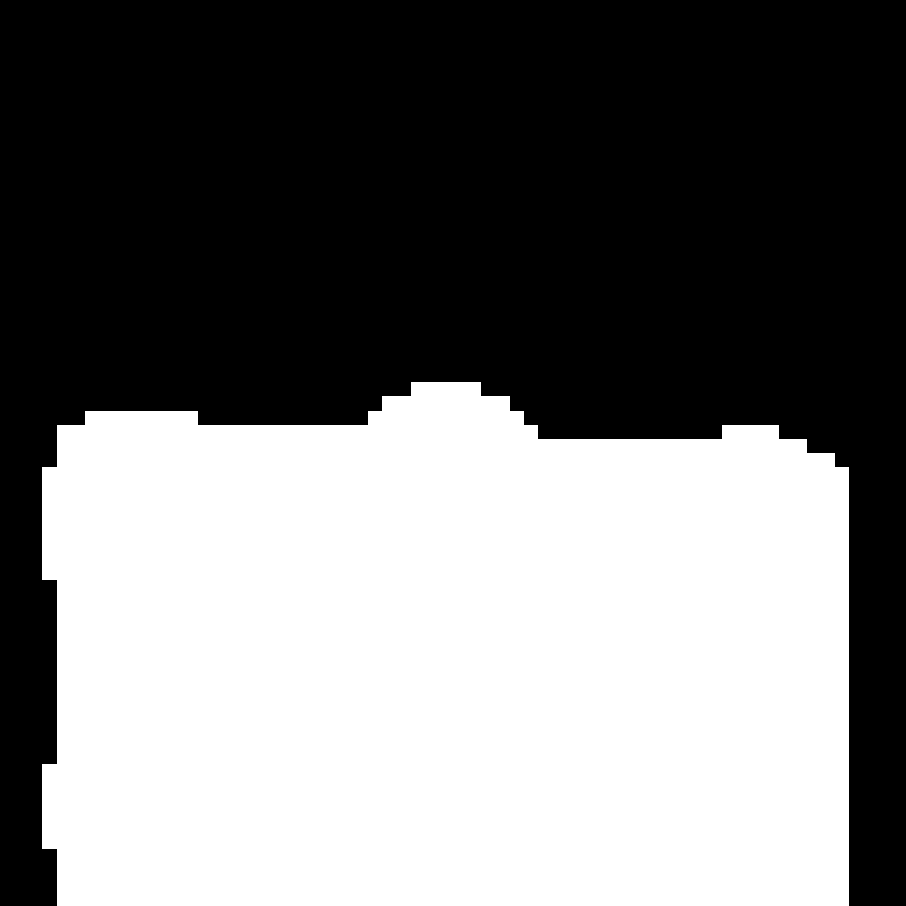}}~
\subfloat[Joint \label{fig:long_segm}]{\includegraphics[width=0.3\textwidth,]{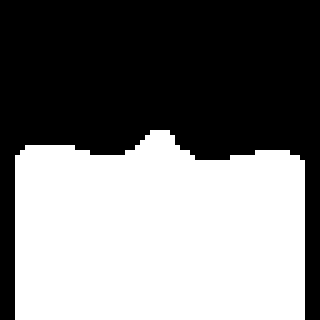}}

\caption{Magnitude reconstructions (top row) and corresponding segmentations (bottom row) for the sequential approach and our joint approach compared to the zero-filling solution. Longitudinal view.} %Results for a bursting bubble from a transversal view (top row) and longitudinal view (bottom row).}
\label{fig:frame4magn_long}
    \end{figure}

    We also present the results for the magnitude and segmentation for the zero-filling solution, sequential approach and joint approach. We can see in Fig.~\ref{fig:frame4magn_trans} and \ref{fig:frame4magn_long} that the joint approach exploits the structure in the data and present more accurate magnitude reconstructions and segmentations. It is clear that, even in this rather simple segmentation problem, the joint approach is able to improve the results of both tasks. This gain is significant in Fig.~\ref{fig:long_segm}. Additionally, the joint magnitudes present very sharp edges distinguishing air and fluid thanks to the segmentation coupling term in the model, which acts as additional prior to reconstruct images exploiting prior knowledge on the region of interest. 
    
    %%magnitude for i=4
    

\section{Conclusion and outlook}
In this work we have presented a joint framework for flow estimation, magnitude reconstruction and segmentation from undersampled velocity-encoded MRI data. After having described the corresponding dynamic inverse problem, we have presented a joint variational model based on a non-convex Bregman iteration. We have demonstrated that by imposing regularity on the individual components (in contrast to the sequential approach), %exploiting the strong correlation in the time-dependent data, 
our joint method achieves accurate estimations of the velocities, as well as an enhanced magnitude reconstruction with sharp edges, thanks to the joint segmentation. Furthermore, we assessed the performance of our joint approach on synthetic and real data. In this context, we have shown that the joint model improves the performances of the different imaging tasks compared to the classical sequential approaches. 

Future work includes the investigation of the full joint temporal and spatial optimisation. By extending the model to the full 4D setting, we believe the performance will be enhanced further, as temporal correlation e.g. in the segmentation can be exploited. The current limitation is the lack of such 4D dataset. Indeed, as described in the acquisition protocol, the velocity data was acquired separately for each spatial component to speed up the acquisition.

\begin{acknowledgement}
VC acknowledges the financial support of the Cambridge Cancer Centre and Cancer Research UK. MB acknowledges the Leverhulme Trust Early Career Fellowship ECF-2016-611, "Learning from mistakes: a supervised feedback-loop for imaging applications". CBS acknowledge support from Leverhulme Trust project "Breaking the non-convexity barrier", EPSRC grant "EP/M00483X/1", EPSRC centre "EP/N014588/1", the Cantab Capital Institute for the Mathematics of Information, and from CHiPS and NoMADS (Horizon 2020 RISE project grant). Moreover, CBS is thankful for support by the Alan Turing Institute.
\end{acknowledgement}

%\input{referenc}
%\clearpage
  \addcontentsline{toc}{section}{Bibliography}
  \bibliographystyle{unsrt}
  \bibliography{velocity} 
  
\section*{Appendix}
In this section we show the full dynamic sequence of a bubble burst event. At time $t=1$ the bubble resting at the air-liquid interface. When the thin liquid film breaks, the bubble burst, causing the formation of an upward and downward jet. The upward jet moves in the empty space left by the bubble and reached its maximum at $t=4$. After that, the jet falls down into the liquid pool, causing a downward jet and some oscillation. At around $t=8$ the liquid motion stops.
\label{app}
\begin{figure}[h!]
   \centering
\subfloat[$t=1$]{\includegraphics[width=0.22\textwidth]{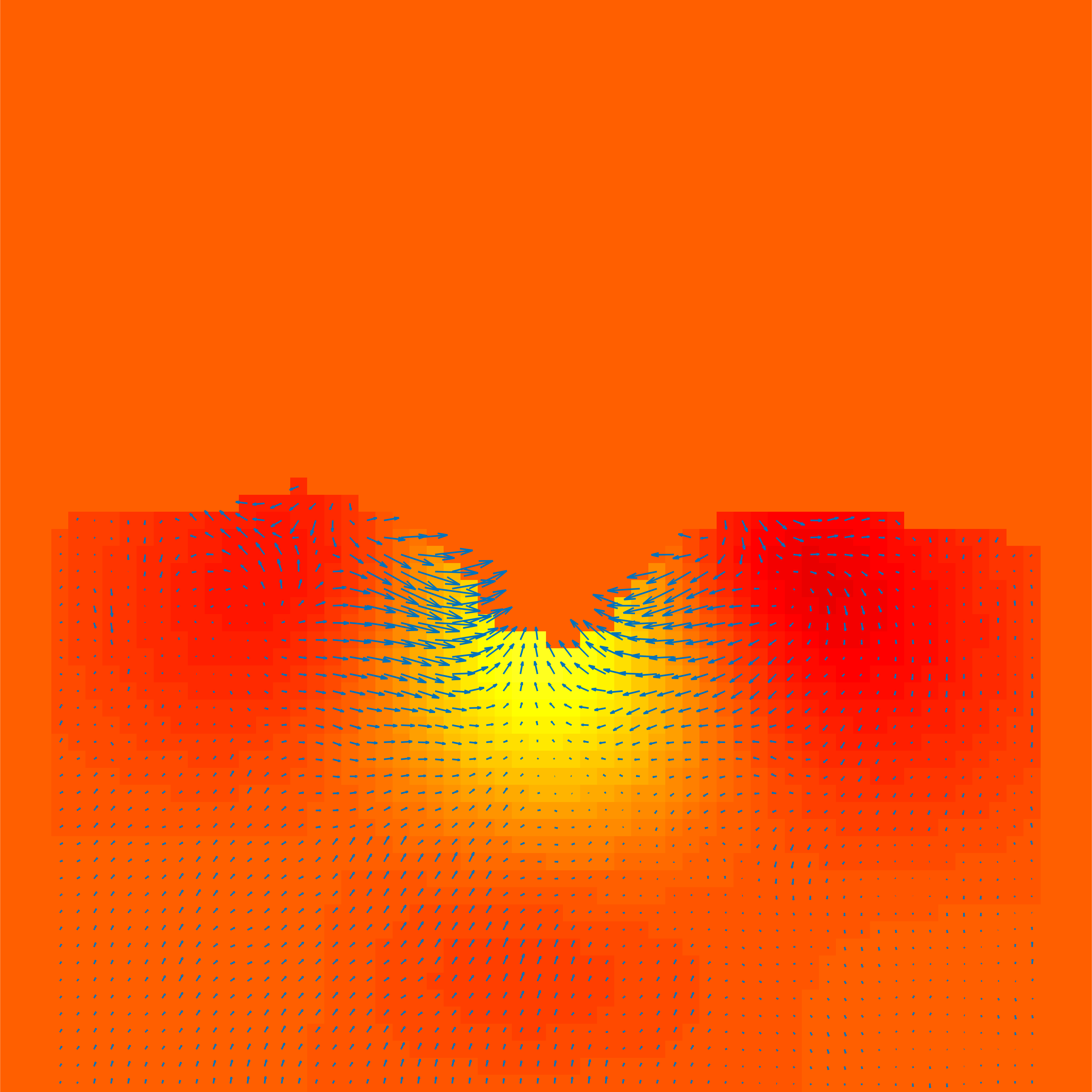}}~
\subfloat[$t=2$]{\includegraphics[width=0.22\textwidth]{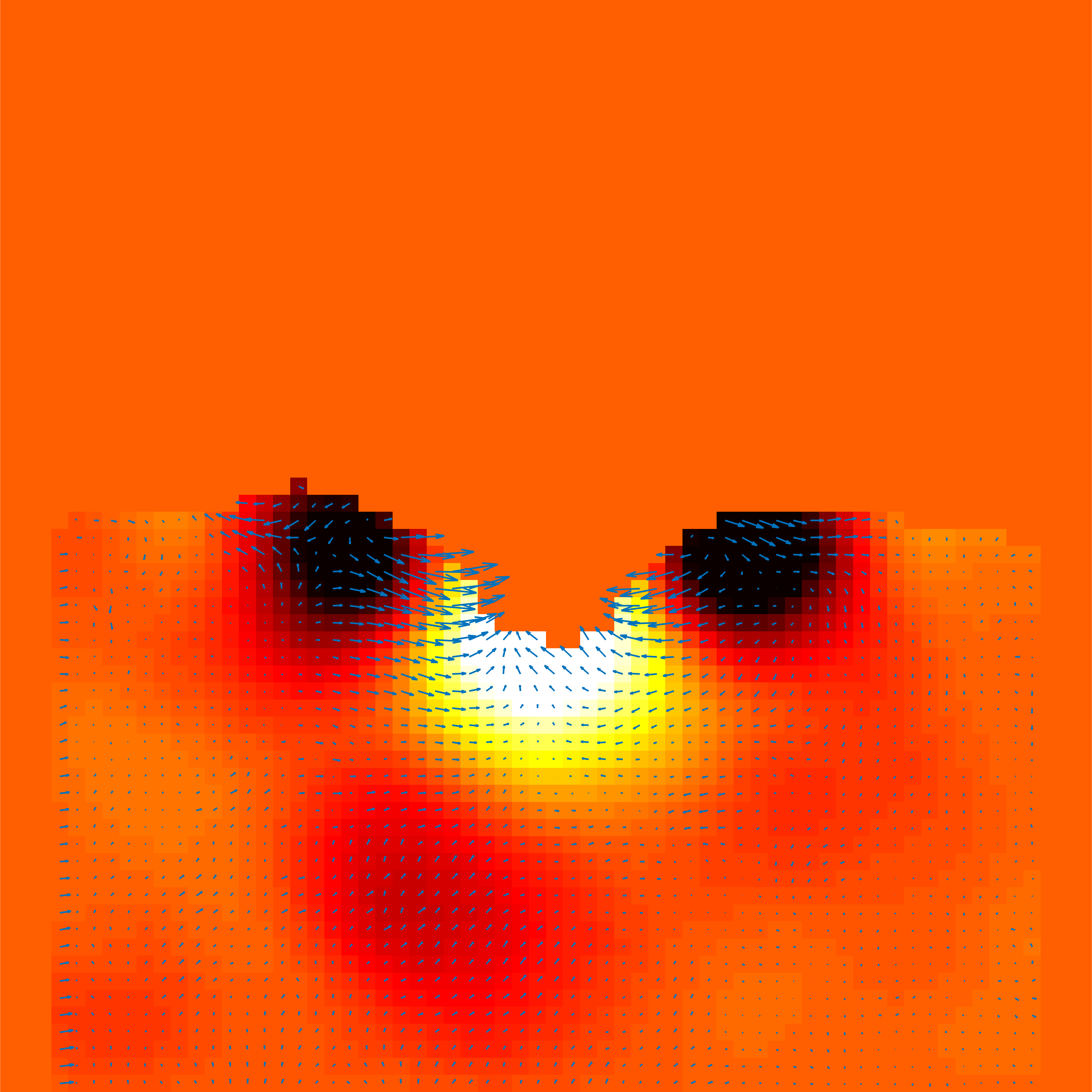}}~
\subfloat[$t=3$]{\includegraphics[width=0.22\textwidth,]{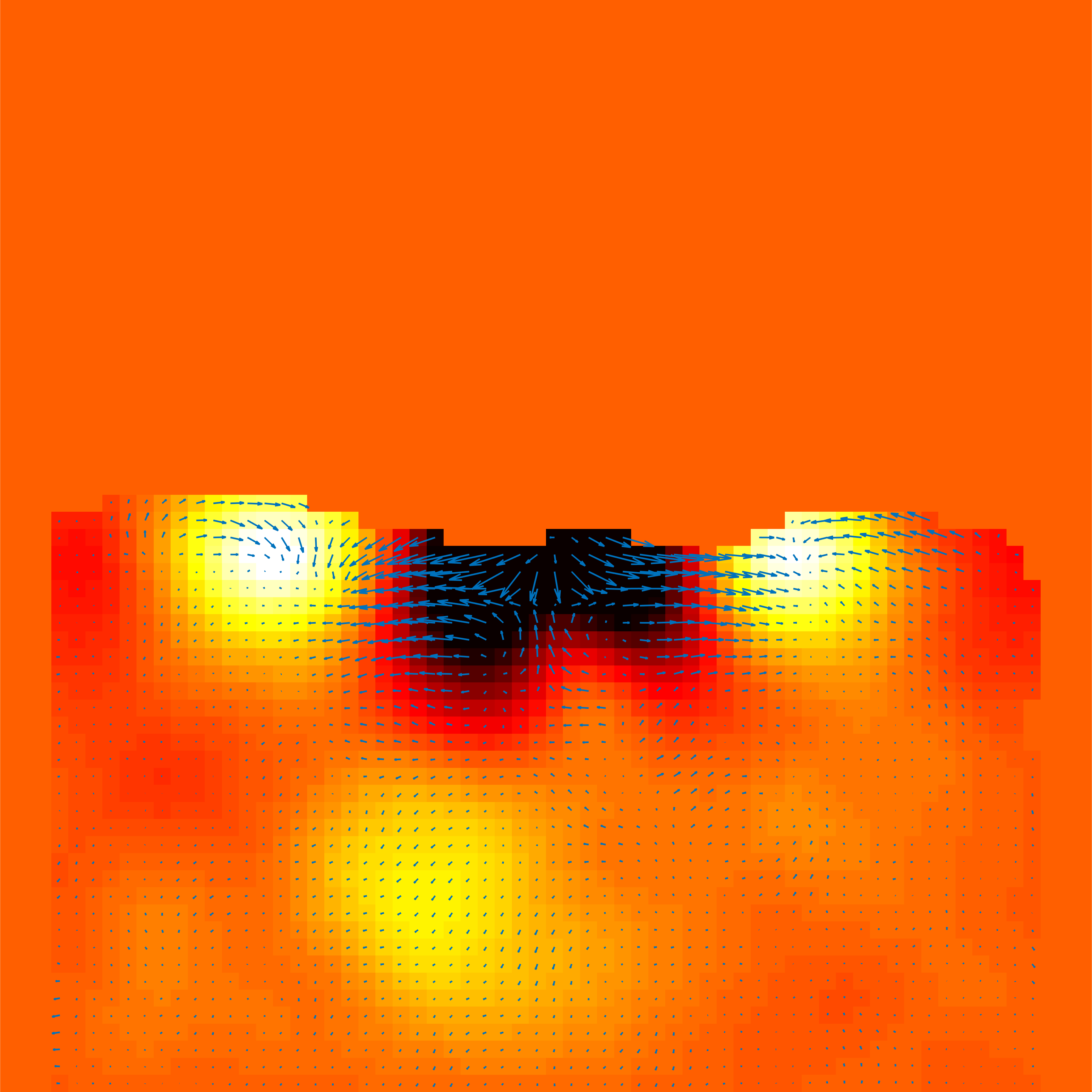}}~
\subfloat[$t=4$]{\includegraphics[width=0.22\textwidth,]{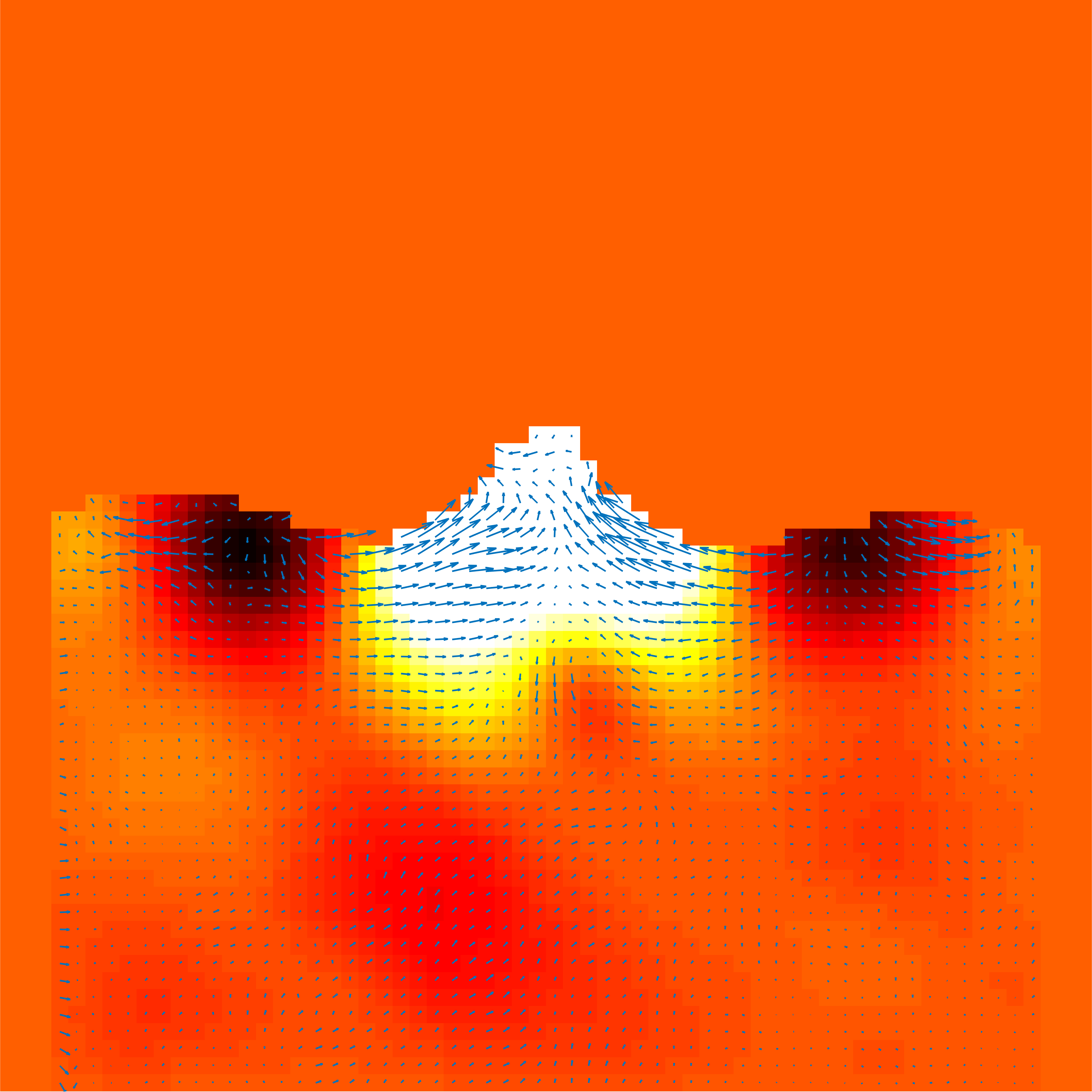}}

   \centering
\subfloat[$t=5$]{\includegraphics[width=0.22\textwidth]{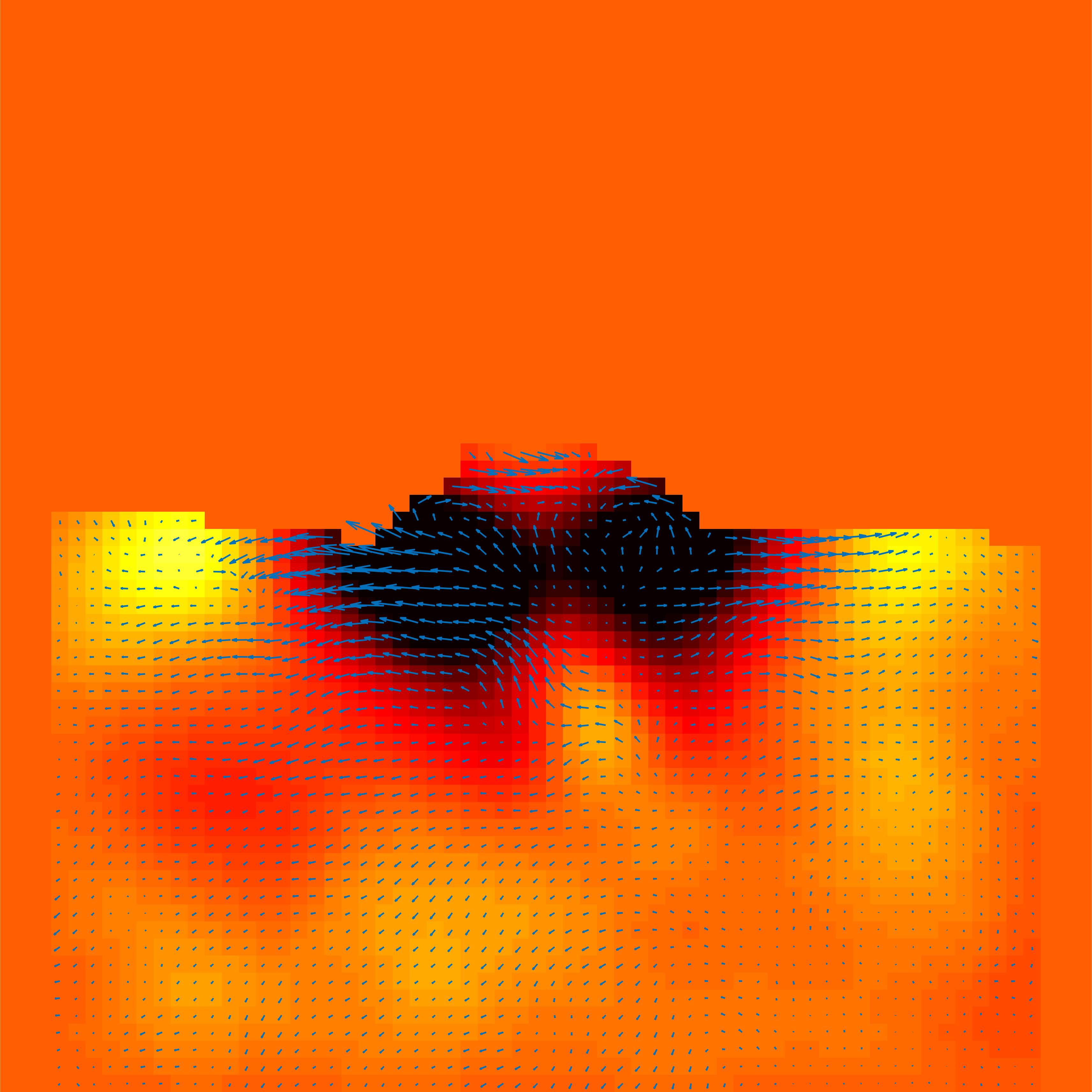}}~
\subfloat[$t=6$]{\includegraphics[width=0.22\textwidth]{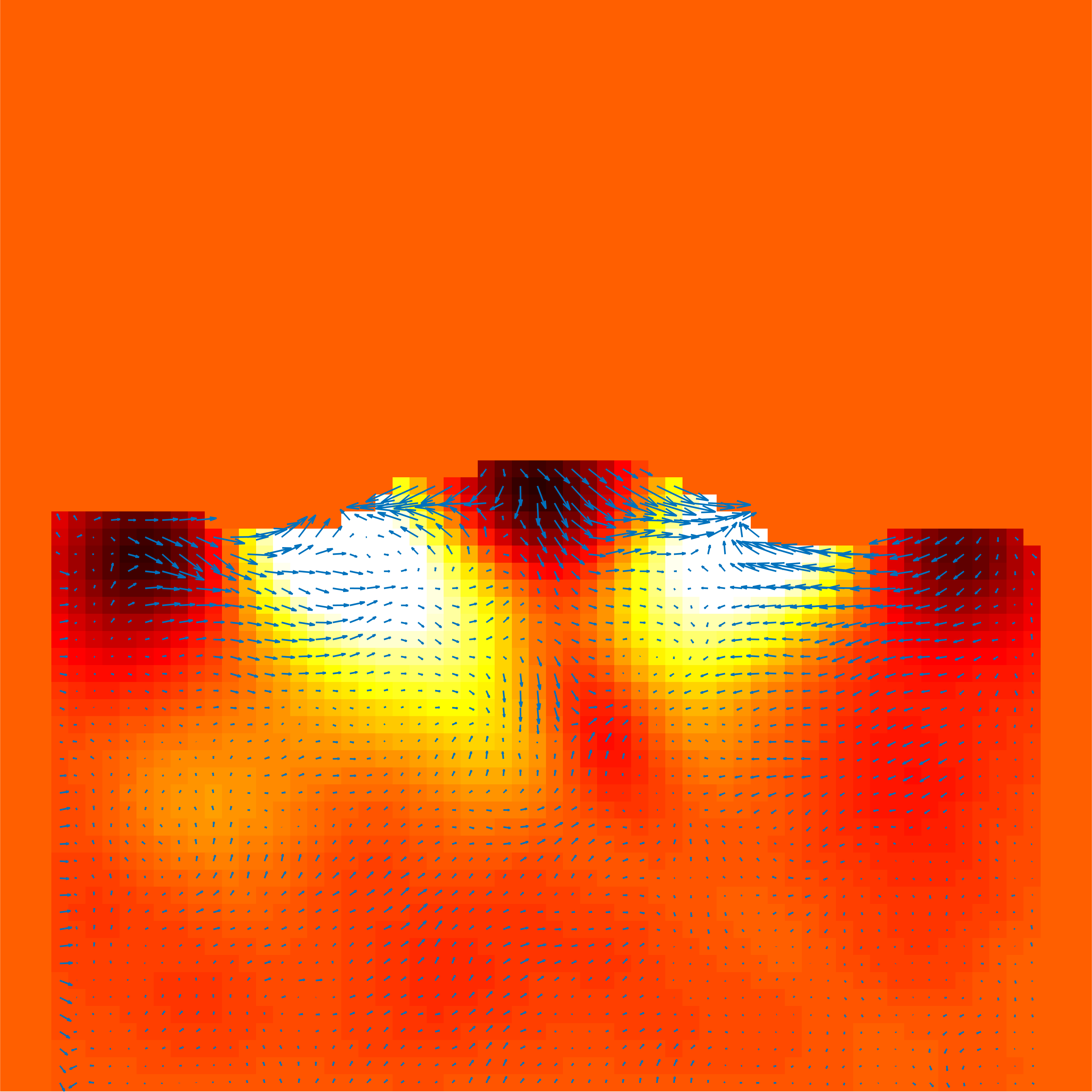}}~
\subfloat[$t=7$]{\includegraphics[width=0.22\textwidth,]{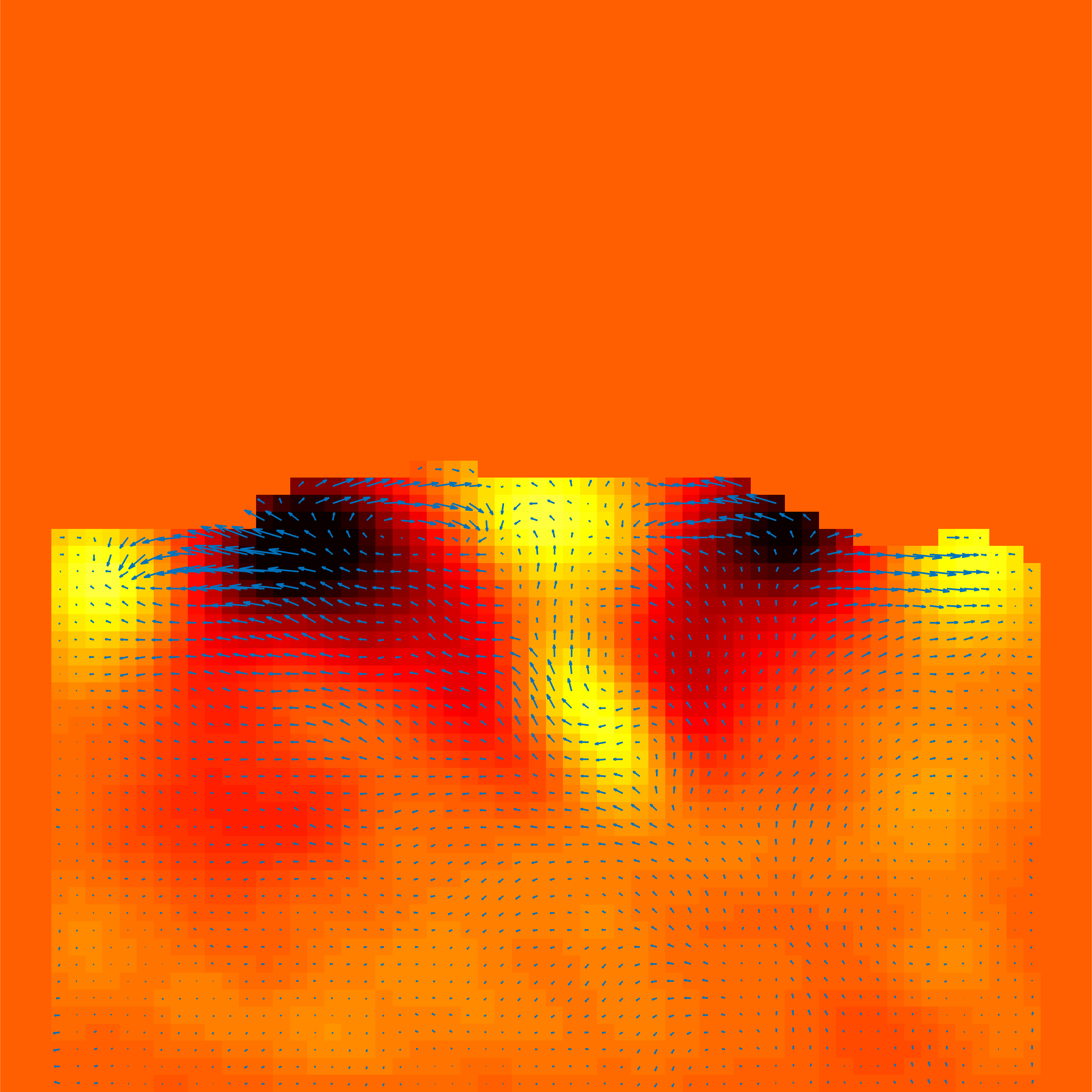}}~
\subfloat[$t=8$]{\includegraphics[width=0.22\textwidth,]{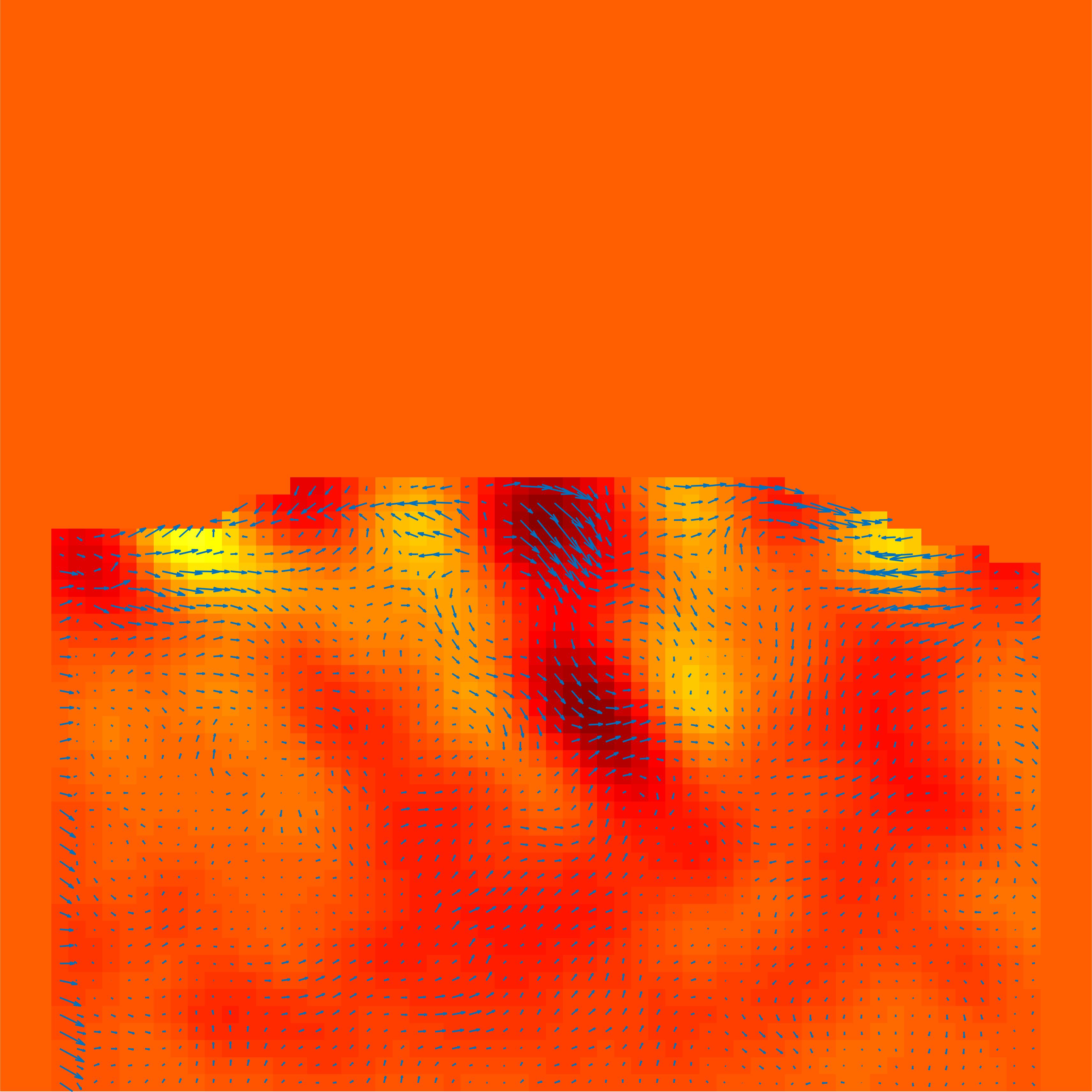}}
\caption{Full time sequence. Longitudinal view. The bubble burst event sees the bubble resting at the interface between liquid and air, before this film is finally broken. The bursting causes an upward jet that moves the liquid at its highest position at $t=4$. Subsequently, the jet drops into a downward jet, causing oscillation in the liquid, until it finally dies out at $t=8$.}
\label{fig:long}
    \end{figure}
    
    \begin{figure}[h!]
   \centering
\subfloat[$t=1$]{\includegraphics[width=0.22\textwidth]{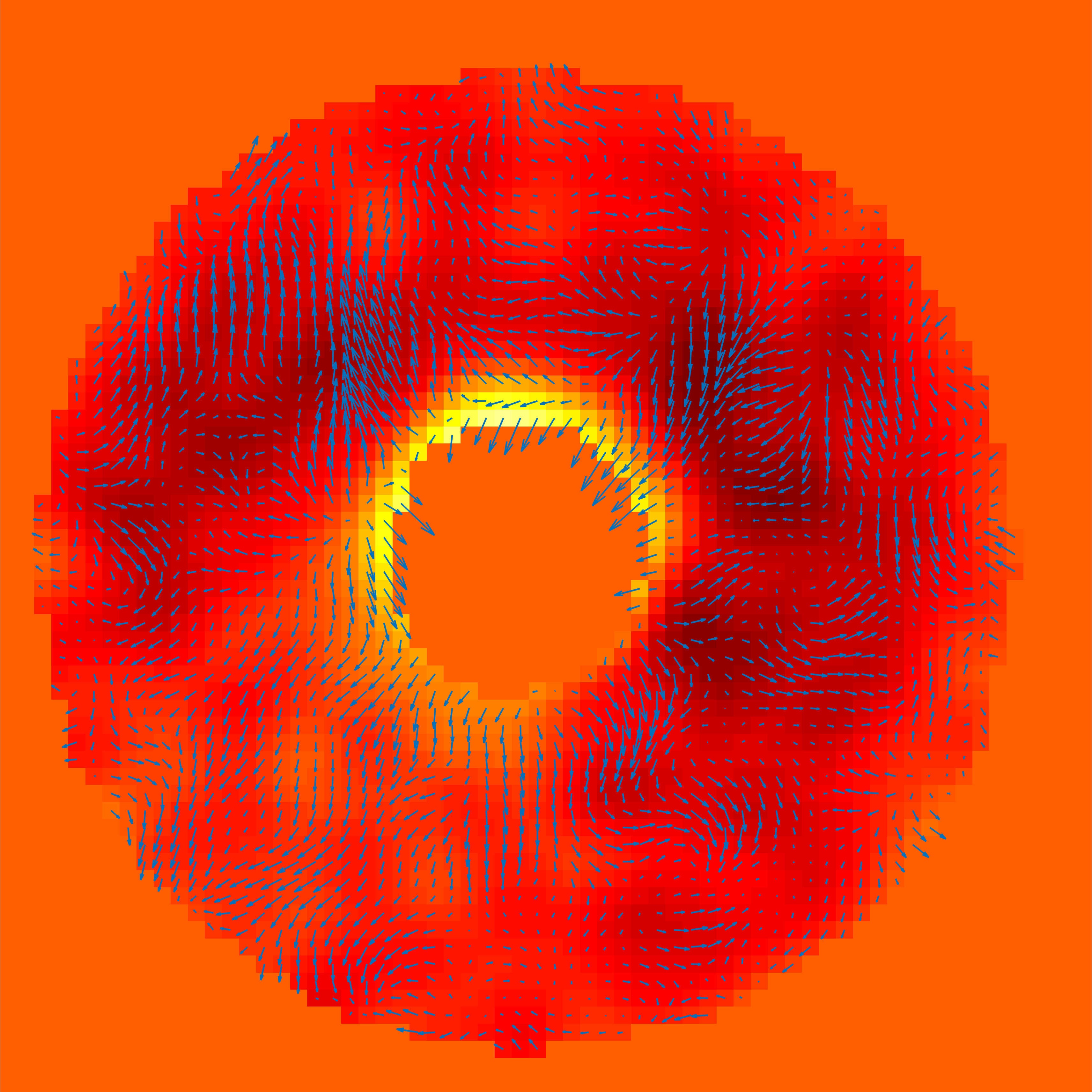}}~
\subfloat[$t=2$]{\includegraphics[width=0.22\textwidth]{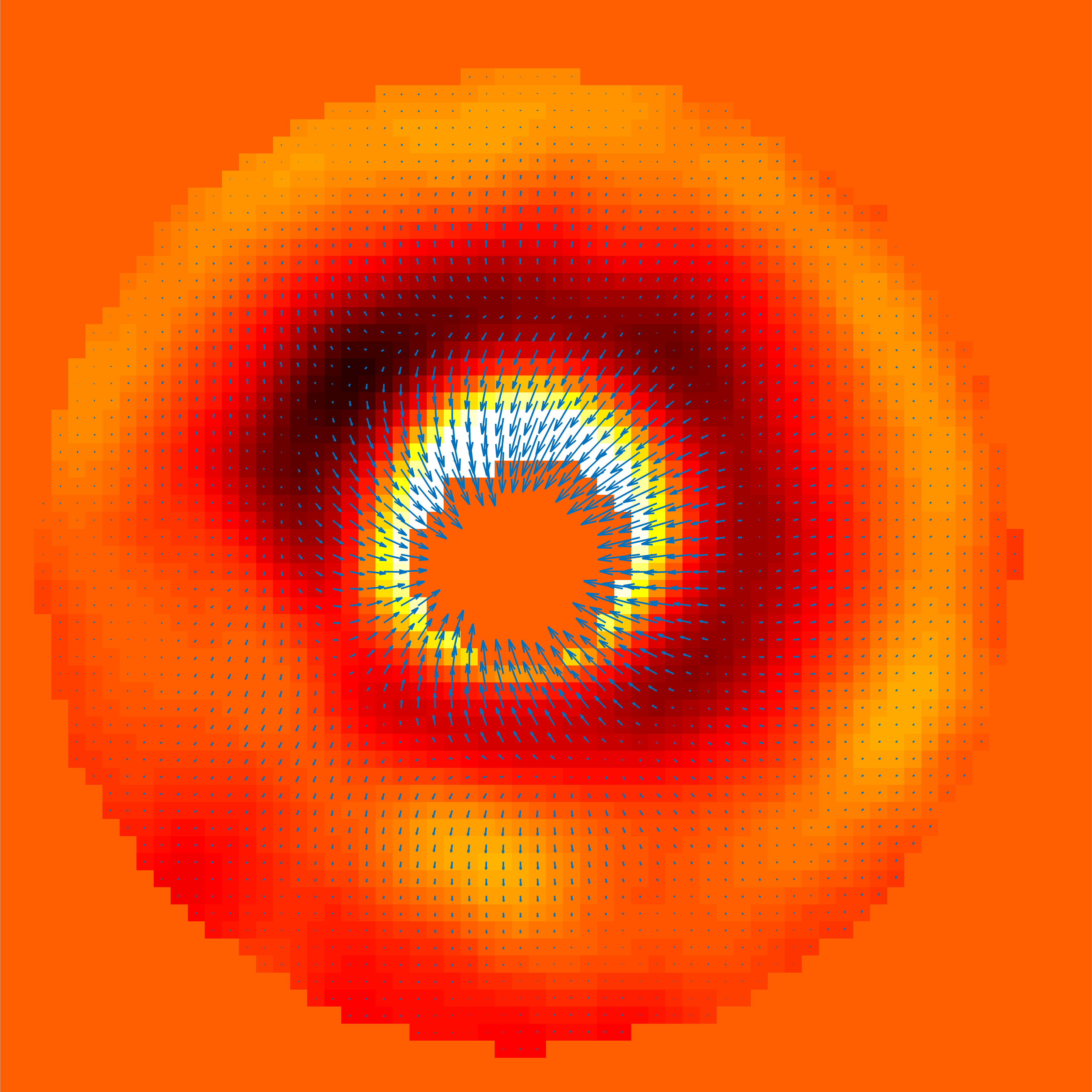}}~
\subfloat[$t=3$]{\includegraphics[width=0.22\textwidth,]{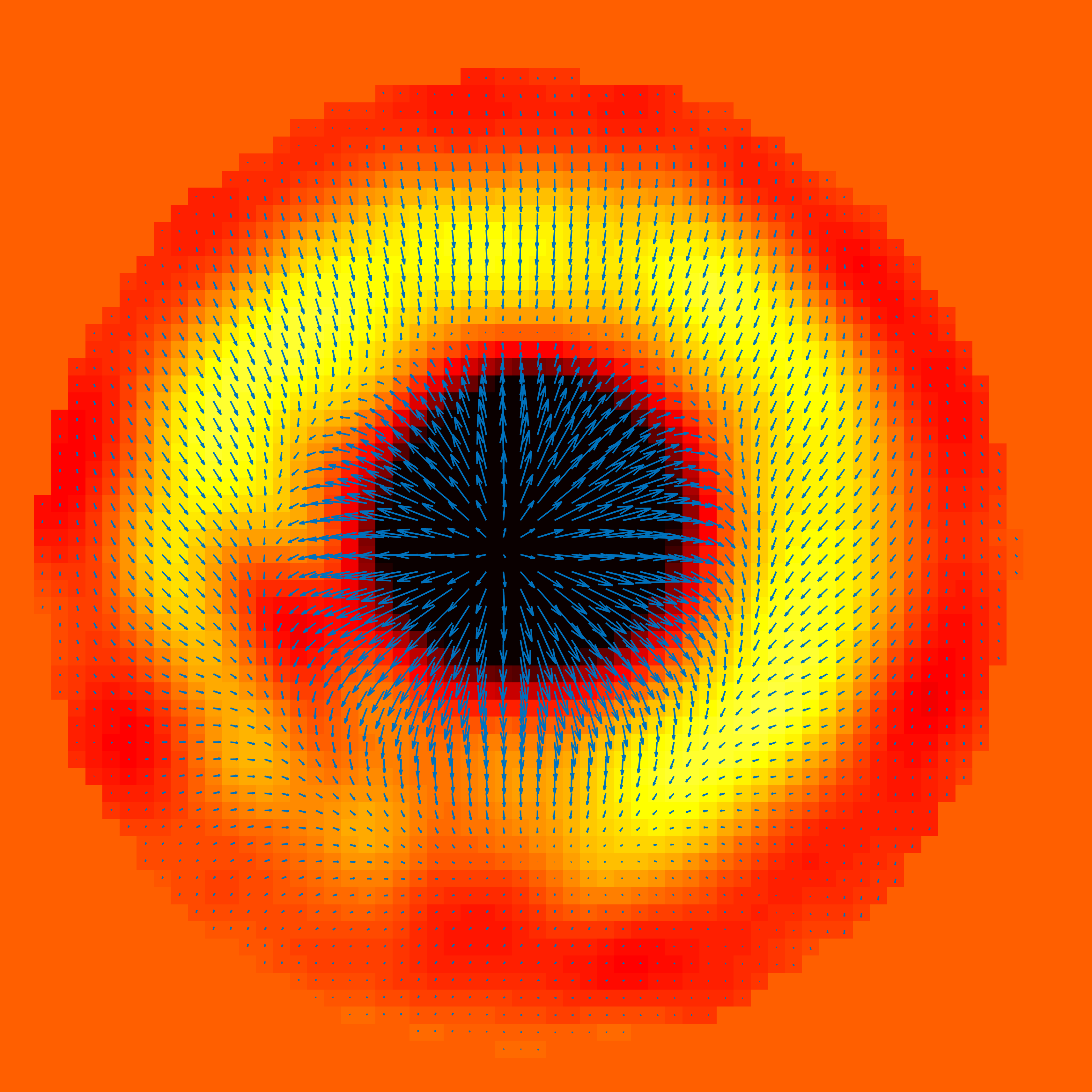}}~
\subfloat[$t=4$]{\includegraphics[width=0.22\textwidth,]{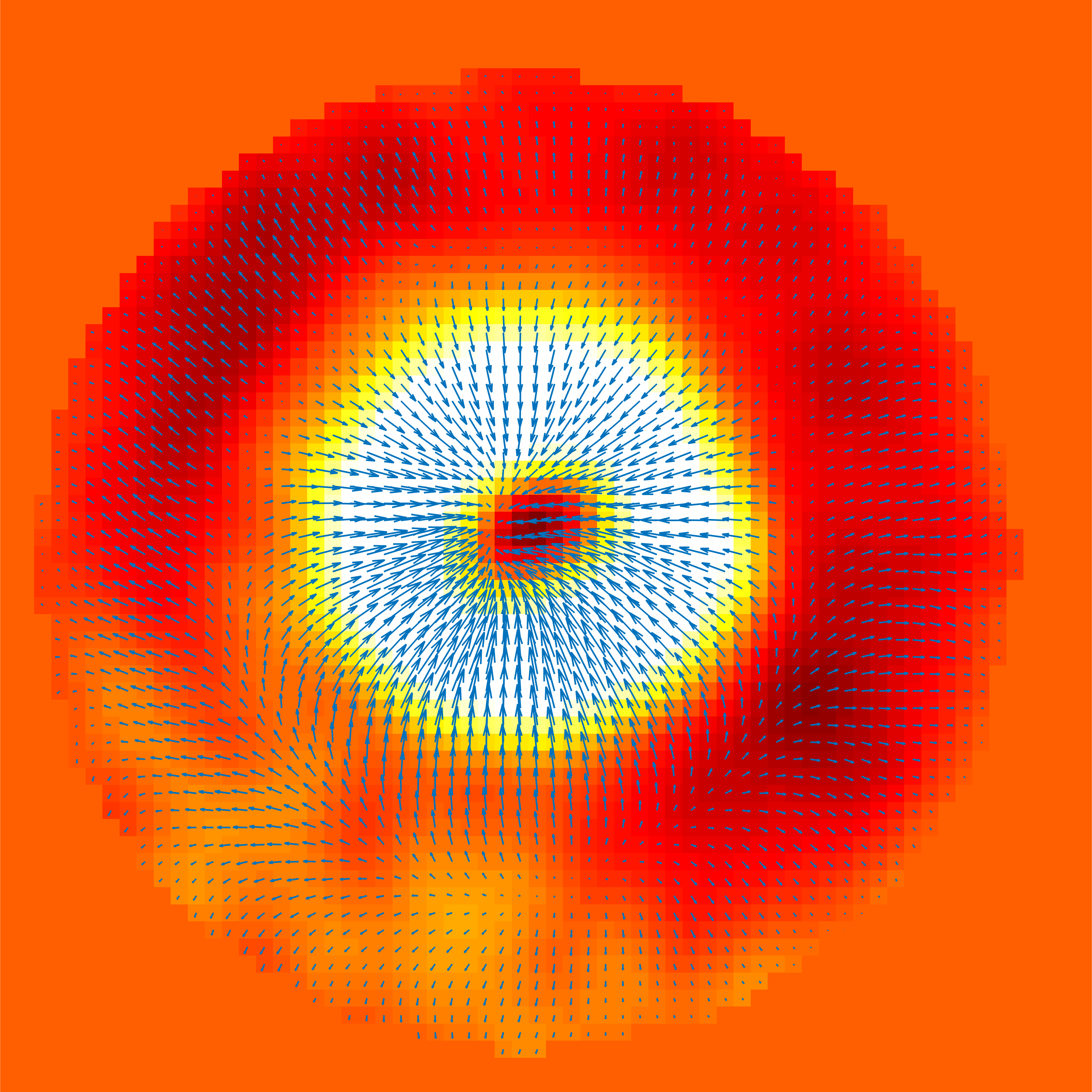}}

   \centering
\subfloat[$t=5$]{\includegraphics[width=0.22\textwidth]{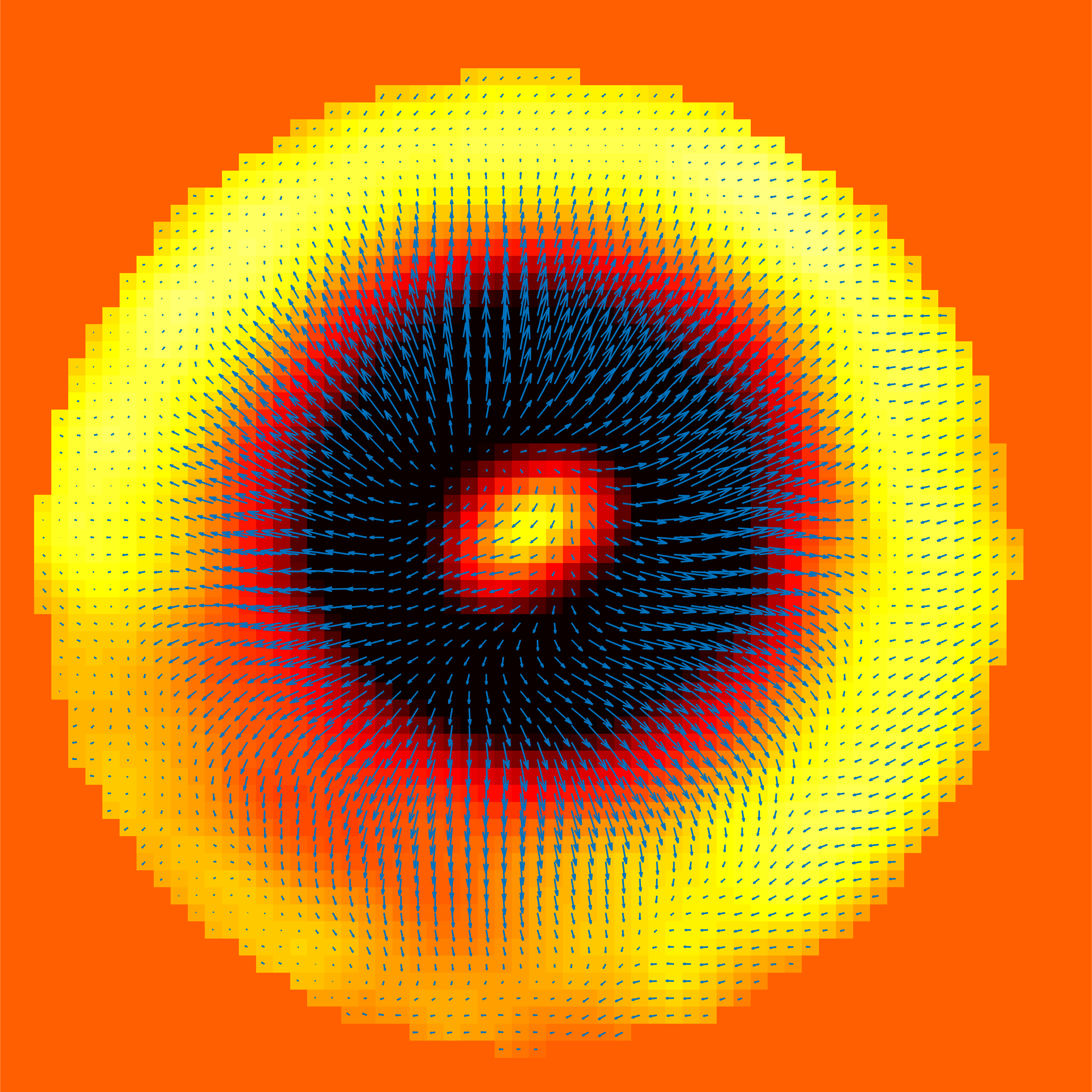}}~
\subfloat[$t=6$]{\includegraphics[width=0.22\textwidth]{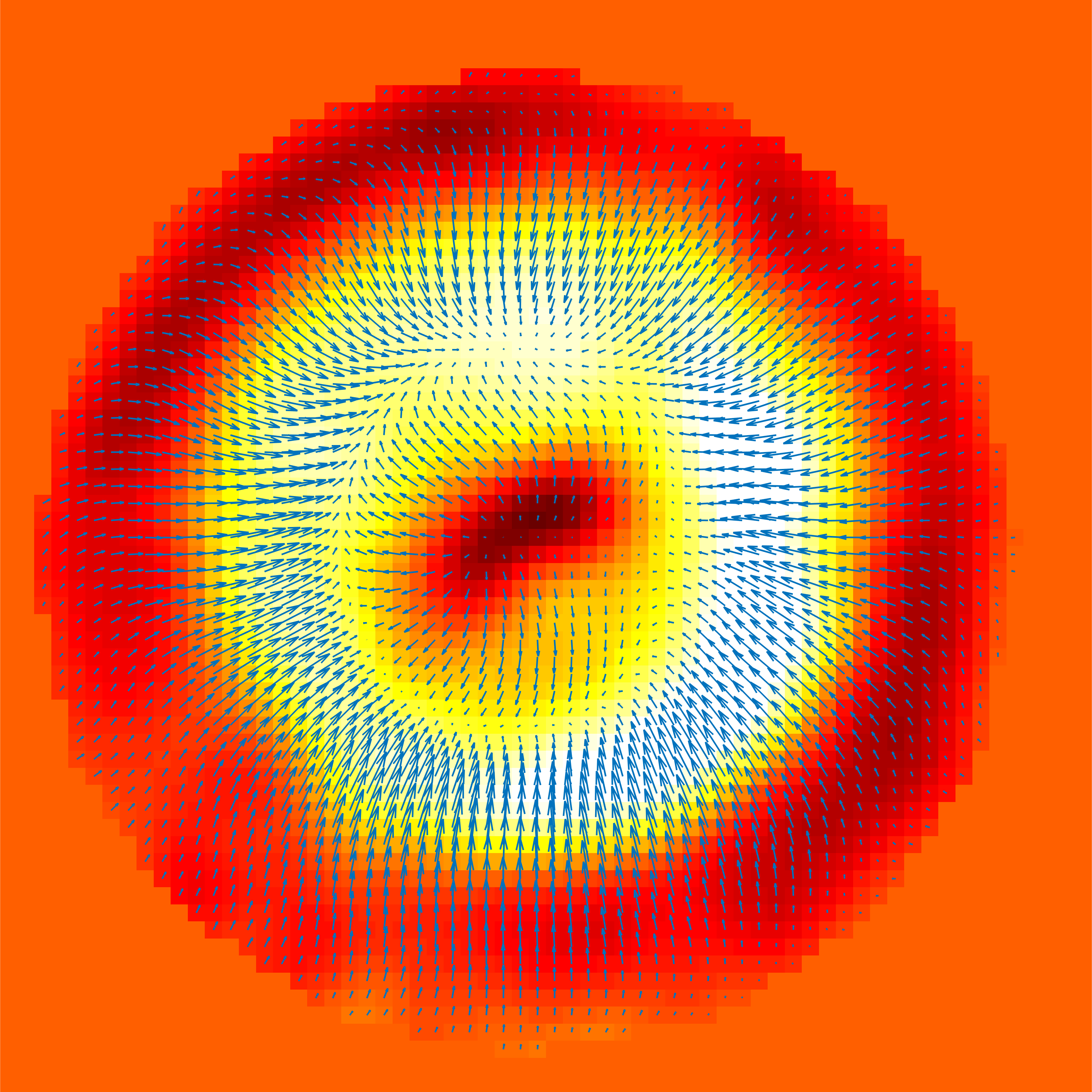}}~
\subfloat[$t=7$]{\includegraphics[width=0.22\textwidth,]{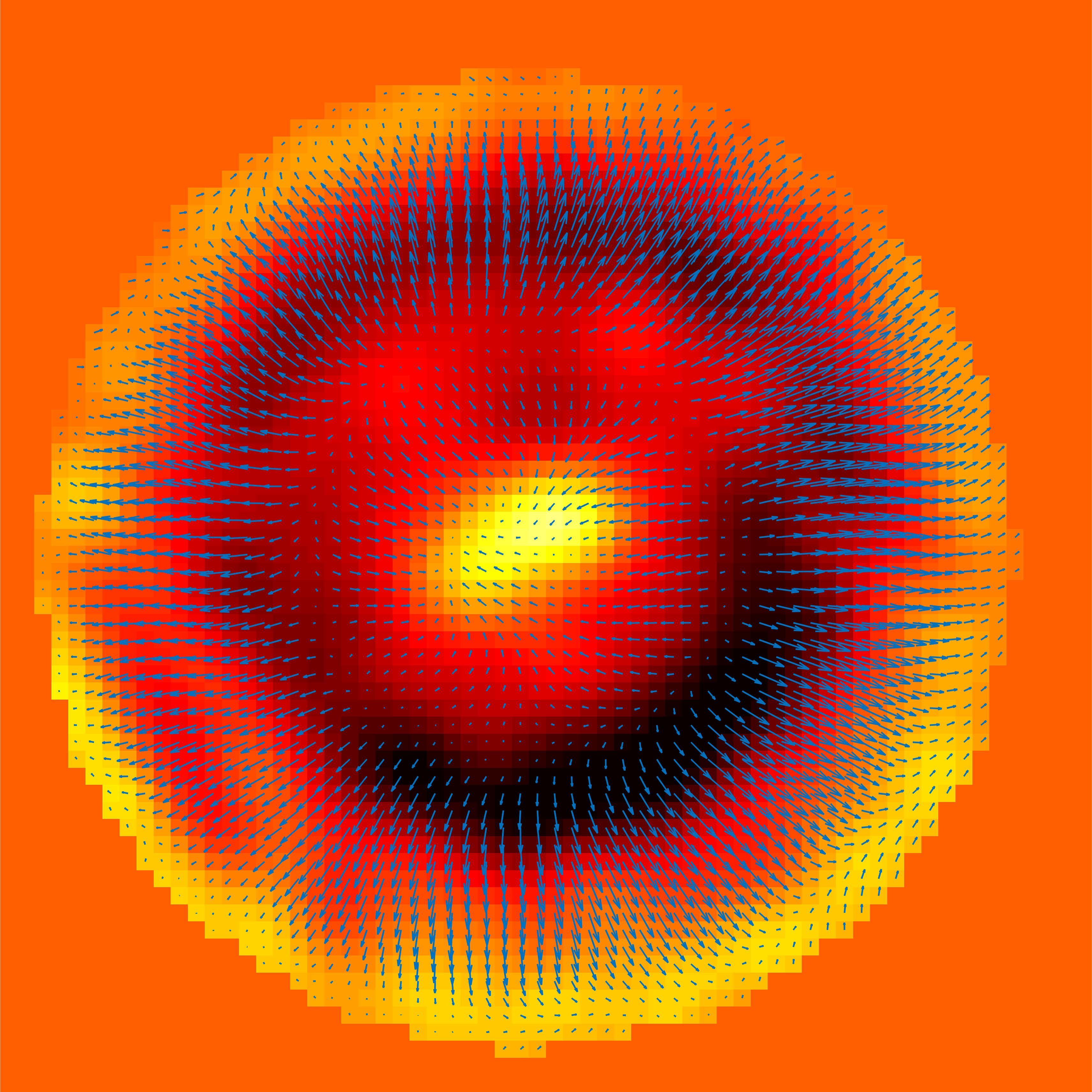}}~
\subfloat[$t=8$]{\includegraphics[width=0.22\textwidth,]{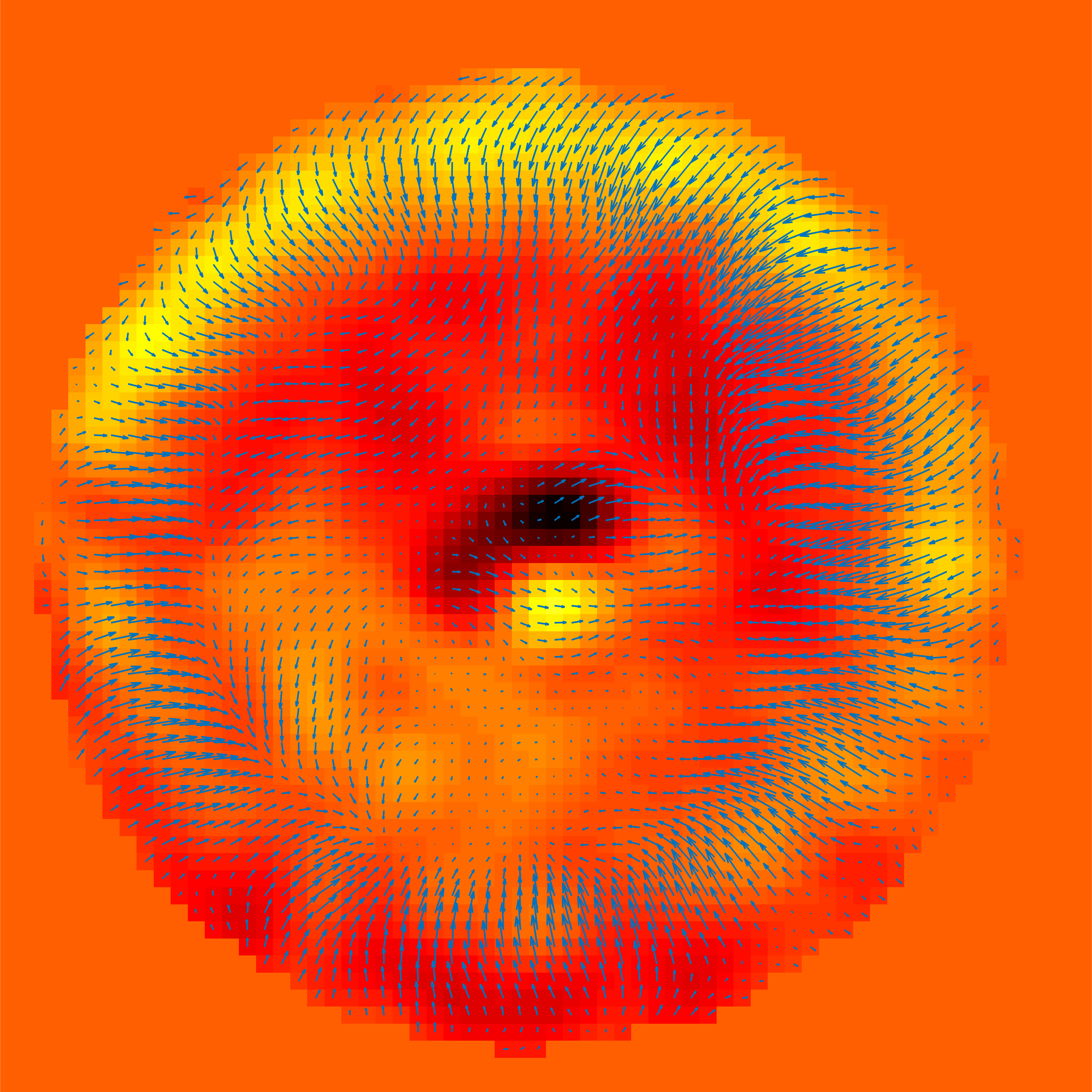}}
\caption{Full time sequence. Transversal view through the middle of the bubble. We can see the bubble burst event and the upward/inward jet caused by the empty space left by the bubble. Subsequently, the jet falls down into the liquid pool causing a downward/outward jet, until it dies out at $t=8$.}
\label{fig:trans}
    \end{figure}
\end{document}